\documentclass[a4paper,fleqn,usenatbib]{mnras}

\usepackage{txfonts}
\usepackage[T1]{fontenc}
\usepackage{ae,aecompl}
\usepackage{pdflscape}
\usepackage{subcaption}
\usepackage{etoolbox}
\usepackage{multirow}
\usepackage{graphicx}	

\usepackage{amssymb}	

\title[The impact of baryonic physics]{Cosmological simulations of the same spiral galaxy: the impact of baryonic physics}

\author[A. Nu\~nez-Casti\~neyra et al.]{
A. Nu\~nez-Casti\~neyra,$^{1,2}$\thanks{E-mail: arturo.nunez@lam.fr (KTS)}
E. Nezri,$^{1}$
J. Devriendt$^{3}$
and R. Teyssier$^{4}$
\\
$^{1}$Aix Marseille Univ, CNRS, CNES, LAM, Marseille, France\\
$^{2}$Aix Marseille Univ, CNRS/IN2P3, CPPM, Marseille, France\\
$^{3}$Sub-department of Astrophysics, University of Oxford, Keble Road, Oxford OX1 3RH, UK\\
$^{4}$Institute for Computational Science, University of Z\"urich,
CH-8057 Z\"urich, Switzerland
}

\date{Accepted XXX. Received YYY; in original form ZZZ}

\pubyear{2015}

\begin{document}

\label{firstpage}
\pagerange{\pageref{firstpage}--\pageref{lastpage}}
\maketitle

\begin{abstract}
	The interplay of star formation and supernova (SN) feedback in galaxy formation is a key element for understanding galaxy evolution. Since these processes occur at small scales, it is necessary to have sub-grid models that recover their evolution and environmental effects at the scales reached by cosmological simulations. In this work, we present the results of the \emph{Mochima} simulation, where we simulate the same spiral galaxy inhabiting a Milky Way (MW) size halo in a cosmological environment changing the sub-grid models for SN feedback and star formation. We test combinations of the Schmidt law and a multi-freefall based star formation with delayed cooling feedback or mechanical feedback. We reach a resolution of 35 pc in a zoom-in box of 36 Mpc. For this, we use the code RAMSES with the implementation of gas turbulence in time and trace the local hydrodynamical features of the star-forming gas. Finally, we compare the galaxies at redshift 0 with global and interstellar medium observations in the MW and local spiral galaxies. 
	The simulations show successful comparisons with observations. Nevertheless, diverse galactic morphologies are obtained from different numerical implementations. We highlight the importance of detailed modelling of the star formation and feedback processes, especially for simulations with a resolution that start to reach scales relevant for molecular cloud physics. Future improvements could alleviate the degeneracies exhibited in our simulated galaxies under different sub-grid models.
\end{abstract}




\section{Introduction}
In a $\Lambda$CDM universe, proto galactic haloes are formed as dark matter gravitationally collapses from initial overdensities. The galaxies are then formed through the subsequent accretion of baryonic gas and dark matter. As the host halo grows, so does its gas content reaching high densities that would locally collapse and form stars. If star formation (SF) were only described by the gravitational collapse of the gas, this process would be faster and more efficient than what is observed \citep{1974ApJ...192L.149Z,2007ApJ...654..304K,2009ApJS..181..321E, 2014PhR...539...49K}. The fact that molecular clouds survive for longer than their associated freefall time suggests that there are other processes involved in star formation as the galaxy forms. Several models have been proposed to explain the inefficiency of star formation, from stellar feedback \citep{2010ApJ...709..191M} and turbulent support \citep{2004RvMP...76..125M,krumholz2005general,hennebelle2011analytical,2011ApJ...730...40P} to dynamical stabilization \citep{2010ApJ...721..975O,2018ApJ...854..100M}, and magnetic fields \citep{2015MNRAS.450.4035F} highlighting the complexity of the interstellar medium (ISM). 

At galactic scales, abundance matching techniques \citep{2010MNRAS.404.1111G,behroozi2010comprehensive, munshi2013reproducing, rodriguez2015stellar, kravtsov2018stellar} give an insight into the relation between the mass of the host dark matter (DM) halo and its baryonic component. From such works, we know that the peak of galactic star formation occurs around Milky Way size haloes, suggesting a boundary between the main processes that dominate over star formation in smaller and bigger haloes than $\approx 10^{12}$ M$_{\odot}$ \citep{1986ApJ...303...39D,1998A&A...331L...1S}.

The non-linearity in the formation and evolution of galaxies make cosmological simulations a powerful tool to compare models and observations. All this by featuring dark matter collapsing into large scale structures and baryonic gas collapsing into stars to form galaxies. To this end, it is necessary to include sub-grid models to describe processes like star formation and feedback that take place at resolutions far below those reached by these simulations. Either with directly coupled hydrodynamics or semi-analytic models, these implementations come as far as to reproduce fundamental general observables like the Kennicutt-Schmidt relation, star formation histories, rotation curves, and stellar to halo mass relation in big volume simulation \citep{2013MNRAS.436.3031V,2014MNRAS.444.1453D,2015MNRAS.446..521S,2015MNRAS.454...83W,2015ARA&A..53...51S}. Regardless of the general success of large volume simulations, high-resolution studies suggest that the current sub-grid implementations might not be enough to fully reproduce galaxy formation \citep{2012MNRAS.423.1726S}, either on the side of feedback \citep{2012MNRAS.421.3488H,2013ApJ...770...25A,kimm2015towards,2017MNRAS.470.3167V,rosdahl2016snap} or the formation of stars coupled to the ISM modeling \citep{2015IAUGA..2257403P,2017MNRAS.466.4826K,2018MNRAS.477.1578H,2019arXiv190611836K}. More sophisticated implementations of the sub-grid processes are needed to better describe galaxy formation and reproduce lower scale observations.

Initially, in numerical simulations of galaxy formation, only SN feedback was used to constrain star formation. The approach was to thermally inject the SN energy into the neigbouring environment \citep{1992ApJ...391..502K}. This technique resulted in very compact and dense galaxies given that the energy was rapidly radiated away without affecting its local environment, with almost no effect on SF \citep{dalla2012simulating}. Along the last 20 years, considerable progress has been achieved in developing models of SN feedback, mainly aiming at reproducing large scale observables \citep{1993MNRAS.265..271N,1997A&A...325..972G}. It has also been proposed in the light of observations of the turbulent nature of the ISM \citep{1974ApJ...192L.149Z,1981MNRAS.194..809L,1987ApJ...319..730S,1992A&A...257..715F,2002A&A...390..307O,2004ApJ...615L..45H} that the supersonic random motions of gas in molecular clouds actually regulates SF.  We now understand that the turbulence has a double purpose when it comes to star formation. Turbulent kinetic energy impedes the gravitational collapse of the molecular cloud on large scales. On smaller scales, the formation of dense filaments through the action of the shocks can form dense cores that serve as star formation sites. Such a system will result in a clumpy star distribution over the spiral arms instead of the incorrect smooth distribution usually observed in simulations\citep{2015MNRAS.450.4035F}.
 
In this paper, we present a comparison of the effect that different sub-grid implementations have on a spiral galaxy in a cosmological environment. We reach a resolution of ~35 pc and store dynamical variables of the gas cells, which allows us to compare the properties of the star-forming gas in our simulations with observations of local star-forming regions. In particular we focus on the star formation implementation used by \citet{2017MNRAS.466.4826K} and \citet{2015IAUGA..2257403P}, the mechanical feedback approach of \citet{kimm2015towards}, the delayed cooling feedback approach by \citet{Teyssier2013} and the ISM turbulent implementation of \citet{2019arXiv190611836K}.
 The paper is organized as follows, in section \ref{sec:thesimulation} we describe the simulation framework and the sub-grid strategies for star formation and SN feedback. In section \ref{sec:results}, we present the results of the simulations and compare them with observations in two main blocks, global galactic properties in section \ref{subsec:Globalprop} and local or small scale properties in section \ref{subsec:Localprop}. Finally, in section \ref{sec:Conclusions} we present our summary and conclusions.

\section{Simulation}\label{sec:thesimulation}
Simulations were run with the Eulerian adaptive mesh refinement (AMR) code RAMSES \citep{Teyssier2002}, to study the impact of different implementations of baryonic physics on a spiral galaxy hosted by a Milky Way size halo. Initial conditions were generated with the MUSIC \citep{hahn2013music} package, generating the primordial density fluctuations at redshift 100 in a periodic box of 36 Mpc containing a $\Lambda$CDM universe. We use as cosmological parameters: H$_0=67.8$ km/s/Mpc for the value of Hubble constant today, $\Omega_{\mathrm{b},0}=0.045$ the baryonic matter density, $\Omega_{\mathrm{m,0}}=0.308$ for matter density and $\Omega_{\Lambda,0}=0.692$ for the vacuum density. We start by evolving only the dark matter content of the box with a uniform resolution up to redshift 0. Once there, the HAST\footnote{writtern by V. Perret and available at   \url{https://bitbucket.org/vperret/hast/wiki/Home}} code is used to select the haloes that fulfil the MW-like halo mass, merger history and environment criteria i.e. M$_h\leq 10^{12}$ M$_{\odot}$ \citep{2012ApJ...759..131B,2012ApJ...761...98K,Mcmillan2016}, no major mergers after redshift 2 and no massive neighbour halo. Then with HAST, we generate the convex hull Lagrangian volume. This volume contains the initial positions of all particles that end up inside 3.5 times the virial radius, $r_{\mathrm{vir}}$, of the final halo (see \citet{2014MNRAS.437.1894O}). After identifying the Lagrangian region new initial conditions are  generated with 5 resolution levels on DM particle mass, starting from the outer box with a local resolution that corresponds to 128$^3$ particles and ending with a resolution of 2048$^3$ particles ( i.e. 11 levels of refinement or $m_{dm} = 1.9 \times 10^5 \mathrm{M}_{\odot}$) inside the Lagrangian volume. This volume's boundaries are redefined, if necessary, to include particles of lower resolution that cross inside $r_{\mathrm{vir}}$ during the halo evolution, this process is known as decontamination. Once the Lagrangian volume is decontaminated, baryons are included in the initial conditions and a full hydrodynamics run of the zoom-in halo is done until redshift 0. Ending in what we call the \emph{Mochima} galaxy, a spiral galaxy with a central bulge, the total stellar and DM mass are comparable to those expected for Milky Way. The analysing tools where developed from the data reading tool UNSIO\footnote{written by Jean-Charles Lambert from the CESAM group at LAM and available at \url{https://projets.lam.fr/projects/unsio} }.

 The primary properties of the five runs are listed in Table \ref{tab:mainnumbers}. In what follows we use the definition of the virial radius as the radius at which the mean density reaches the critical density of the universe $\rho_{\mathrm{crit}}$ times the so-called virial overdensity $\Delta_{\mathrm{crit}}= 18\pi+82x-39x^2$ where x is defined as $ x=(\Omega_m/(\Omega_m+a^3\Omega_{\Lambda})-1)$ \citep{1998ApJ...495...80B}.

\subsection{Baryonic physics } \label{sec:Bar}
Two determining processes of galaxy formation are star formation (SF) and stellar feedback. These processes occur at scales that are beyond current available computational resources for galaxy formation simulations, especially in cosmological environments. AMR techniques focus computing time by adaptively dividing space resolution of regions defined by some refining criterion. In our case, this happens when the dark matter or baryonic mass in a cell surpasses a given threshold value. Depending on the object to be simulated, a minimal cell size is necessary to resolve the corresponding characteristic scales, the radius or the scale height of the disc, for example. A compromise needs to be achieved to constrain computing times. Therefore, it is necessary to impose a maximum refining level. The Milky Way's thin disc is reported to have a scale height of $\sim 300$ pc \citep{Mcmillan2016}. We chose to limit our refinement strategy to reach a resolution (minimal cell size $\Delta x$) of 35 pc that allows to resolve a disc scale height similar to that of the MW thin disc with $\sim 8$ cells.    

Even with such resolutions, the scales of the ISM physics remain below the smallest cell of our grid with molecular clouds size ranging from a few to hundreds of parsecs. Here is where a sub-grid numerical prescription is needed to have an effective description of the physics contributing to galaxy formation simulations. Such models have been around for more than two decades \citep{1992ApJ...399..331C,1992ApJ...399L.109K}. Considering the technological advances in both the computing resources and ISM observations, it is now necessary to expand such simple models to include gas dynamics.

 A full theory of star formation remains to be developed. Nevertheless, we now understand the role of local gas turbulence as a competitor to the gravitational collapse of gas in the ISM \citep{2004RvMP...76..125M,2007ARA&A..45..565M}. We highlight the impact of baryonic physics modelling on a simulated spiral galaxy in a cosmological context. We now describe our main sub-grid prescriptions for turbulence, star formation, and SN feedback.

\subsubsection{Turbulence}
Using the Navier-Stokes equation to describe a fluid's turbulence numerically requires reaching microscopic resolutions. Such scales are out of reach of galaxy formation simulations in cosmological environments. An alternative approach is to relate large scale motions of the fluid, such as turbulence to the mean properties of the flow. This methodology was modelled in the 60s by \citet{1963MWRv...91...99S} and are now called Large Eddy Simulations (LES). In astrophysics, the most often used sub-grid scale (SGS) models have been based on numerical dissipation. It is then assumed that large scale (above resolution) dynamics are more or less independent of the sub-resolution fluctuations and therefore the latter can be smoothed out \citep{1984JCoPh..54..174C}.

 LES models were later introduced in astrophysics to describe supernova combustions \citep{2006A&A...450..283S,2005CTM.....9..693S}, supersonic turbulent flows \citep{2011A&A...528A.106S}, and finally, it has been implemented in the context of star formation for isolated spiral galaxy simulations \citep{2018ApJ...861....4S} and galaxy formation in cosmological environments \citep{2019arXiv190611836K,2020arXiv200303368K}.
 
 Relating large scale motions of the flow with its mean properties, the density field is decomposed in the density averaged over volume, smoothed at resolution scale, $\bar{\rho}$  and the fluctuation $\rho'$. In parallel, the temperature and velocity field are averaged using a mass-weighted average (Favre average) which are denoted as $\tilde{T}$ and $\tilde{v}$ which leads to
\begin{equation}
\rho = \bar{\rho}+\rho', \;\;\;\; T=\tilde{T}+T'', \;\;\;\; v= \tilde{v}+v''
\end{equation}

\noindent fluctuations over the Favre average are denoted with a double prime. Finally, the turbulent kinetic energy that will be stored as a passive scalar is defined as

\begin{equation}
K_T = \frac{1}{2}\overline{ \rho v''^2 } = \frac{1}{2}\bar{\rho}\sigma_{3D}^2
\end{equation}

\noindent where the 1D velocity dissipation $\sigma$ can be related to its three dimensional counterpart and the turbulent kinetic energy as $\sigma^2 = \sigma_{3D}^2 /3 = (2/3) K_T $ for more details we refer the reader to \citet{2011A&A...528A.106S} and \citet{2014nmat.book.....S}.  We use the LES implementation done by \citet{2019arXiv190611836K}, where a modified version of the Euler equation is not used. Only an extra equation for the turbulent kinetic energy is used to account for advection and work of turbulent pressure as in  \citet{2014nmat.book.....S}, and \citet{2018ApJ...861....4S}

\begin{equation}
\frac{\partial}{\partial t}K_T + \frac{\partial}{\partial x_j}(K_T \tilde{v_j})+P_T\frac{\partial \tilde{v_j}}{\partial x_j} = C_T- D_T\;\;,
\end{equation}
\noindent where the turbulent kinetic energy is related to the turbulent pressure by $P_T = 2/3 K_T$ and the \textit{creation term} has the following form 
\begin{equation}
	C_T = 2\mu_T \sum_{ij}\left[
		\frac{1}{2}\left(\frac{\partial \tilde{v_i}}{\partial x_j}+
		\frac{\partial \tilde{v_j}}{\partial x_i}
		\right) - \frac{1}{3}\left( 
		\mathbf{\nabla} \cdot \tilde{\mathbf{v}} \right)\delta_{ij}
	\right]\;\;.
\end{equation}

\noindent Here, the \textit{destruction term} is responsible for the dissipation of the turbulence in the sub-grid turbulent cascade and is modelled as

\begin{equation}
	D_T = \frac{K_t}{\tau_{\mathrm{diss}}}\;\;.
\end{equation}

\noindent This model has two important parameters, the turbulent viscosity $\mu_T$ and the dissipation time scale $\tau_{\mathrm{diss}}$, which are related to the cells size by

\begin{equation}
	\mu_T = \bar{\rho}\Delta x \sigma\;\;\;\mathrm{and}\;\;\; \tau_{\mathrm{diss}}= \frac{\Delta x}{\sigma} 
\end{equation}

	Previous implementations of thermo-turbulent star formation sub-grid models consider an in-situ calculation of the turbulent velocity dispersion \citep{2015IAUGA..2257403P,2017MNRAS.470..224T,2018MNRAS.478.5607T,2018MNRAS.477.1578H}(which would be equivalent to considering both terms \textit{creation} and \textit{destruction} to be equal). In our case, this model is used to estimate the turbulent velocity dispersion over time using the density and velocity fields without modifying the hydrodynamic solver. The obtained velocity dispersion will play a key role in the turbulent star formation model described below.

\begin{table*}
\caption{\label{tab:mainnumbers} Global values of the six runs of the Mochima galaxy, one dark matter only and five hydro runs. From left to right are the tag of the galaxy, the protostellar feedback parameter $\epsilon$, the total halo mass, the total stellar mass inside $r_{\mathrm{vir}}$, the stellar mass inside $0.2\;r_{\mathrm{vir}}$, the virial radius, maximum resolution of the mesh, dark matter particle mass in the zoom region and the minimum mass of a star particle present in the simulation.  }
\begin{tabular}{|c|c|c|c|c|c|c|c|c|c|}
\hline
                            & \multirow{2}{*}{Tag}      & \multirow{2}{*}{$\epsilon$} & $M_{\mathrm{Halo}}$    & $M_{\mathrm{stars}}$    & $M_{\mathrm{stars},0.2}$ & $r_{\mathrm{vir}}$ & $\Delta x$             & $m_{\mathrm{dm}}$      & $m^{\star}_{\mathrm{min}}$ \\
                            &                           &                             & $(10^{12}\;M_{\odot})$ & $(10^{10}\; M_{\odot})$ & $(10^{10}\; M_{\odot})$  & (kpc)              & (pc)                   & $(10^5 \; M_{\odot})$  & $(10^5 \; M_{\odot})$      \\ \hline
Dark matter only            & -                         & -                           & 1.129                  & -                       & -                        & 275.9              & 140.5                  & 2.279                  & -                          \\ \hline
Schmidt law+Delayed Cooling & KSlaw-DCool               & -                           & 0.923                  & 3.128                   & 3.066                    & 260.7              & \multirow{5}{*}{35.13} & \multirow{5}{*}{1.947} & \multirow{5}{*}{0.1568}    \\ \cline{1-7}
Multi-ff KM+Delayed Cooling & Mff$\epsilon_{009}$-DCool & 0.09                        & 0.950                  & 7.436                   & 7.321                    & 266.6              &                        &                        &                            \\ \cline{1-7}
Multi-ff KM+Delayed Cooling & Mff$\epsilon_{100}$-DCool & 1.00                         & 0.917                  & 3.701                   & 3.618                    & 266.5              &                        &                        &                            \\ \cline{1-7}
Multi-ff KM+Mechanical FB   & Mff$\epsilon_{009}$-MecFB & 0.09                        & 0.979                  & 10.58                   & 10.10                    & 272.5              &                        &                        &                            \\ \cline{1-7}
Multi-ff KM+Mechanical FB   & Mff$\epsilon_{100}$-MecFB & 1.00                         & 0.938                  & 8.037                   & 7.597                    & 271.3              &                        &                        &                            \\ \cline{1-7}
\end{tabular}
\end{table*}

 \subsubsection{Star formation}
The first star-formation (SF) approach we use is motivated by the Schmidt law \citep{kennicutt1998global}, and consist of keeping a constant SF efficiency over the full simulation. The SF rate  is computed as

\begin{equation}\label{eq:starformation}
\dot{\rho}=\epsilon_{\mathrm{ff}} \frac{\rho_{\mathrm{gas}}}{t_{\mathrm{ff}}} \;\;\;\; \rho_{\mathrm{gas}}> n_{\star}
\end{equation}

\noindent where $\rho_{\mathrm{gas}}$ is the gas density of the cell and $\epsilon_{\mathrm{ff}}$  is the SF efficiency per free-fall time $t_{\mathrm{ff}} = \sqrt{3\pi/32 G \rho_{\mathrm{gas}}}$. This means that $100 \epsilon_{\mathrm{ff}} \%$ of the gas mass in the cell will be turned into stars as long as the cell is denser than the threshold density $n_{\star}$. The threshold density can be calculated by requiring the Jeans length to be larger than four times the smallest cell in the simulation \citep{2014MNRAS.444.2837R}. This leads to $n_{\star}=19.182 $ H/cc. In previous works, this calculation was also used to set a temperature floor for the gas evolution in order to avoid numerical fragmentation. We choose to use this calculation only to compute the threshold density for the control run using Schmidt law SF, and therefore depart from the ``polytropic pressure floor'' by not setting a temperature floor for the gas. It has been argued that the numerical fragmentation that the temperature floor approach aims to avoid might be instead natural gas fragmentation and should not be avoided \citep{2008ApJ...680.1083R}. 

We choose the value of the fixed efficiency to be $\epsilon_{\mathrm{ff}}=0.09$, almost one order of magnitude bigger than the efficiency chosen for similar simulations \citep{2014MNRAS.444.2837R, Mollitor2014} due to the difference in mesh resolution. In our control run this efficiency remains constant regardless of the gas dynamics and forces the SF to be related exclusively to the cell's density.

 The idea of a constant SF efficiency is challenged by works on small-scale numerical simulations \citep{2011ApJ...730...40P,2012ApJ...761..156F} and ISM observations \citep{2011ApJ...729..133M,2016ApJ...833..229L,2018ApJ...861L..18U} that suggest that $\epsilon_{\mathrm{ff}}$ depends on the physical properties of the gas. Therefore we also adopt a thermo-turbulent approach for SF similar to ones used in \citet{2017MNRAS.466.4826K,2017MNRAS.470..224T,2018MNRAS.478.5607T,2019arXiv190611836K}. The full details of the method are beyond the scope of the present document but we give a short description for the sake of completeness.

This SF approach, that we label multi-freefall or multi-ff, following \cite{2012ApJ...761..156F}\footnote{In particular we use the formulation that uses the definitions from \citet{2007ApJ...654..304K} and \citet{2012ApJ...745...69K} which leads to the label KM in some of the figures.}, is based on the assumption that a log-normal distribution yields to a good description of the probability distribution function (PDF) for the gas density of a star-forming cloud. Once this is established, $\epsilon_{\mathrm{ff}}$ can be estimated by integrating the cloud PDF (weighted by a freefall time factor) from a threshold density $\rho_{\mathrm{crit}}$ up to infinity. Given that the freefall time depends on the density, this factor should be inside the integral. The solution to the integral in equations (7) or (34) in \cite{2012ApJ...761..156F} is then their equation (41) that has the following form
\begin{equation}\label{eq:efficiency}
\epsilon_{\mathrm{ff}} = \frac{\epsilon}{2\phi_t}
\exp\left(\frac{3}{8} \sigma_s^2 \right)
\left[
 1 + \mathrm{erf}\left( \frac{\sigma_s^2 - s_{\mathrm{crit}}}{\sqrt{2\sigma_s^2}}\right)\right]
\end{equation}

\noindent where the logarithmic density contrast $s = \ln (\rho/\rho_0)$, the mean gas density is $\rho_0$, and the variance of $s$ is $\sigma_s^2 = ln(1+b^2\mathcal{M}^2)$, where $\mathcal{M}$ is the Mach number. We use the turbulent forcing parameter as $b=0.4$ assuming a mixture of solenoidal and compressive modes for turbulence. The only free parameter of this model is the protostellar feedback (PSFB) parameter $\epsilon$ \cite{2011A&A...528A.106S}. This parameter aims to account for feedback processes that occur at the moment of the molecular cloud collapse, when a fraction $(1-\epsilon)$ of the gas is expected to be blown away by winds, jets and outflows \citep{1993ApJ...410..218W,2000prpl.conf..759K,2007prpl.conf..277P,2011ApJ...729...72P,2011MNRAS.417.1054S,2012ApJ...761..156F}. The expelled gas is  then re-injected into the ISM, while the remaining fraction $\epsilon\leq 1$ falls into the protostellar core contributing to the mass of the future star. We use two different extreme values for this parameter in order to bracket its effect in the simulated galaxy, we have chosen to use $\epsilon = 0.09$ and $\epsilon = 1$. For the critical logarithmic density contrast we adopt the definition of \cite{krumholz2005general} 
\begin{equation}
s_{\mathrm{crit}} = \ln\left( \frac{\pi^2}{5}\phi_x^2 \alpha_{\mathrm{vir}}
\mathcal{M}^2 
\right) 
\end{equation}

\noindent where the virial parameter is defined as $\alpha_{\mathrm{vir}}=2E_{\mathrm{kin}}/|E_{\mathrm{grav}}|$ and the rms Mach number $\mathcal{M}=\sigma/c_s$ is built in terms of the velocity dispersion of the gas cell, $\sigma$, and the sound speed in the cell, $c_s$. The empirical parameters $\phi_t = 0.49$ and $\phi_x = 0.19$ are meant to account for uncertainties in the model. 

In both cases, the Schmidt law and the multi-ff schemes, once a gas cell has passed all the constraints and is allowed to form stars, the star particle has N times the mass of the minimal stellar mass, $m^{\star}_{\mathrm{min}}$. The minimal stellar mass corresponds to the baryonic resolution of the simulation. The value of N is computed following a stochastic model by \citet{2006A&A...445....1R} where N is computed using a Poisson distribution with a mean $\lambda = (\epsilon_{\mathrm{ff}} \rho_{\mathrm{cell}} \Delta x^3 /m^{\star}_{\mathrm{min}} )(\Delta t/ t^{\mathrm{o}}_{\mathrm{ff}})$, where $\Delta t$ is the time step of the simulation and $t^{\mathrm{o}}_{\mathrm{ff}}$ is the infall time of a spherical distribution of mass with density $\rho_{\mathrm{cell}}$.

\subsubsection{SN feedback}

In this study, we examine two different SN feedback models and their impact on the evolution of our simulated galaxy. We use the Chabrier initial mass function \citep{2005astro.ph..9798C} where it is assumed that 31$\%$ ($\eta_{\mathrm{SN}=0.313}$) of the stars are heavy stars ($m_{\star}>8 \mathrm{M}_{\odot}$), and that 5$\%$ of the mass of these stars contribute to the metal content of the cell. The SN feedback will start after a time $t_{\mathrm{sne}}$ from the birth of a star particle. The energy injection will correspond to the amount of heavy stars that are contained in the star particle. Note that one star particle represent a group of stars and not a singular realization. In this simulations only type II supernovae are considered. 

First, we use as a control model the more or less ubiquitous Delayed Cooling method, namely its AMR implementation from \citet{Teyssier2013}. This model aims to account for astrophysical non-thermal processes known to occur in SN explosion sites. Such processes affect the dynamics of the propagation of the shock wave below the usual simulation resolution. The local effects of such processes compete with the gas cooling as they return energy to the gas but these contributions decrease with time. The non-thermal energy $e_{\mathrm{SN}}$ evolves as follows
\begin{equation}
\frac{D e_{\mathrm{SN}}}{D t} = \frac{\dot{E}_{\mathrm{inj}}}{\rho} - \frac{e_{\mathrm{SN}}}{t_{\mathrm{diss}}}
\end{equation} 

\noindent meaning that it is driven by the injected SN energy, $\dot{E}_{\mathrm{inj}}$, and damped in the dissipation time, $t_{\mathrm{diss}}$. In practice, the non-thermal pressure is added to the total gas pressure to avoid modifying the hydrodynamical solver. Cooling is neglected while the non-thermal pressure is greater than the thermal pressure and reactivated when they reach comparable magnitudes \citep{Teyssier2013}. For this model the feedback starts after $t_{\mathrm{sne}}=10$~Myr from the birth of the star particle. Following \citet{dubois2015black}, the dissipative time-scale, $t_{\mathrm{diss}}$,  in this approach is determined by the choice of $\eta_{\mathrm{SN}}$, $\epsilon_{\mathrm{ff}}$, $\Delta x$ and $n_{\star}$ 

\begin{equation}
\begin{array}{ll}

t_{\mathrm{diss}}\simeq 
0.82 \left( \frac{\eta_{\mathrm{SN}}}{0.1}\right)^{-1/3}
\times\;\;
\left( \frac{\epsilon_{\mathrm{ff}}}{0.01} \right)^{-1/3}
\times\;\; \\ \\ \;\;\;\;
\left( \frac{N_{\mathrm{cell}} \Delta x}{4 \times 10 \mathrm{pc}} \right)^{2/3}
\times\;\; 
\left(\frac{n_{\star}}{X_H 200 \mathrm{cm}^{-3}}\right)^{-1/6} \mathrm{Myr} 
\end{array}
\end{equation}

\noindent where $X_H=0.76$ is the hydrogen abundance and we take $N_{\mathrm{cell}}=4$ as it is the number of cells where we choose to resolve the Jeans length for the calculations of $n_{\star}$. It can be argued that this model while efficient at galactic scales is not describing the actual physical processes that occur during SN explosions.

Sub-grid models describing the different stages of the SN explosion have been introduced for SPH \citep{2014MNRAS.445..581H}, and more recently for AMR simulations \citep{2014ApJ...788..121K}, we use the so-called mechanical feedback model as described in \citet{kimm2015towards} and study how its effects on a spiral galaxy compare to the effects of the above described delayed cooling method.

In the mechanical feedback approach, the input momentum for the SN event is calculated according to the phases of the Sedov-Taylor  explosion. The main quantity of this model is the ratio between the total swept mass, $M_{\mathrm{swept}}$, and the ejected mass $M_{\mathrm{ej}}$ and is denoted as:
\begin{equation}
\chi = d M_{\mathrm{swept}} / dM_{\mathrm{ej}}
\end{equation}
\noindent where

\begin{equation}
dM_{\mathrm{ej}} = (1-\beta_{\mathrm{sn}}) M_{\mathrm{ej}} /N_{\mathrm{nbor}}
\end{equation}
\noindent and
\begin{equation}
d M_{\mathrm{swept}} = \rho_{\mathrm{nbor}}\left( \frac{\Delta x}{2}\right)^3
+ \frac{(1-\beta_{\mathrm{sn}}) \rho_{\mathrm{host}} \Delta x^3}{N_{\mathrm{nbor}}}
+dM_{\mathrm{ej}}\;\; .
\end{equation}

\noindent Here $\rho_{\mathrm{host}}$ is the cell density, we use $N_{\mathrm{nbor}}=48$ as the number of neighbouring cells (see figure 15 of \citet{2014ApJ...788..121K}). The mass fraction of the summed ejected mass and the mass inside the cell that will stay in the host cell after the SN explosion, is determined by $\beta_{\mathrm{sn}}=4/52$. This value is chosen to attempt an even distribution of the gas mass between the host and the neighbour cells when they are not on the same refining level.

Starting with the free-expansion phase with an available conserved momentum of $\sim 4.5 \times 10^4$km s$^{-1}$ M$_{\odot}$, the momentum increases as more mass is swept by the shock, giving place to the adiabatic so-called Sedov-Taylor phase once the swept mass is comparable to the ejecta mass. As a result, the outward momentum scales as the square root of the total shell mass until the cooling phase starts. In this third stage, the adiabatic expansion ends due to the efficiency of the radiative losses, usually consisting in a very brief period before the start of the last stage, the snowplough phase. 
   In the mechanical feedback implementation, the snowplough phase has a momentum described as \citep{1998ApJ...500..342B,1998ApJ...500...95T,2015ApJ...802...99K,2015MNRAS.448.3248G}.

\begin{equation}
p_{\mathrm{SN,snow}} \approx 3 \times 10^5 \;
\mathrm{km \;s}^{-1} \; \mathrm{M}_{\odot}\;
E_{51}^{16/17}n_{\mathrm{H}}^{-2/17}\;
Z'^{-0.14},
\end{equation}

\noindent where $E_{51}$ is the SN energy in units of 10$^{51}$ erg, $n_{\mathrm{H}}$ is the hydrogen number density and $Z'$ is the metallicity in solar units. The mass ratio that would trigger the transition to the snowplough phase is then 
\begin{equation}
\chi_{\mathrm{tr}} = 69.58\;E_{51}^{-2/17}n_{\mathrm{H}}^{-4/17}
Z'^{-0.28},
\end{equation}
\noindent and the injected momentum evolved as
\begin{equation}
p_{\mathrm{SN}} =
\left\{
    \begin{array}{ll}
        p_{\mathrm{SN,\,ad}} = \sqrt{2\chi M_{\mathrm{ej}} f_e E_{\mathrm{SN}}} & \mbox{if } (\chi < \chi_{\mathrm{tr}}) \\
        p_{\mathrm{SN,\,snow}} & \mbox{if }(\chi \geq \chi_{\mathrm{tr}})
    \end{array}
\right.
\end{equation}

\noindent where to ensure a smooth transition between both regimes the factor $f_e = 1 - \frac{\chi-1}{3(\chi_{\mathrm{tr}}-1)}$ is used. Note that this implementation might still be dependent on resolution and could result in weak feedback, to try to correct for this we have boosted the number of SN per stellar particle by a factor four, therefore enhancing the effect of the overall SN event in the simulation. \parfillskip=0pt\par
\begin{landscape}
\begin{figure}
\centering 
\includegraphics[width=.85\linewidth]{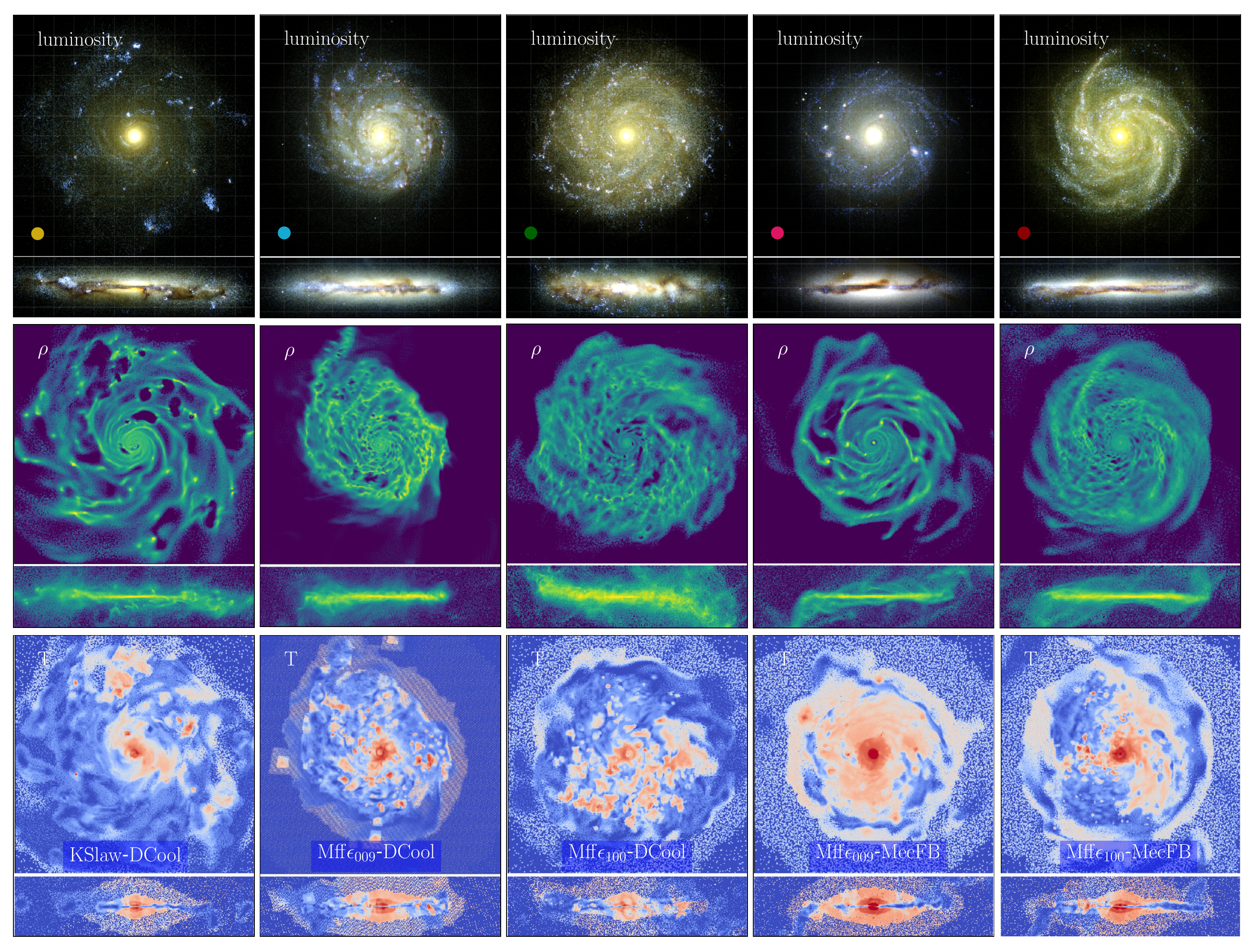}
	\caption{\label{fig:maps} Face-on and edge-on view  of the different simulations at $z=0$. From left to right: KSlaw-DCool, Mff$\epsilon_{009}$-DCool, Mff$\epsilon_{100}$-DCool, Mff$\epsilon_{009}$-MecFB and Mff$\epsilon_{100}$-MecFB. From top to bottom: integrated luminosity along the line of sight of all the stars in SDSS bands including dust obscuration simulated using SKIRT, mid-plane slices of density $\rho$, and temperature T. The grid in the luminosity maps is of 5 kpc and all maps keep the same size of 50 kpc except the left column that have a side of 60 kpc.}
\end{figure}
\end{landscape}

\begin{figure*}
  \centering
  \begin{subfigure}[b]{0.45\textwidth}
    \includegraphics[width=\textwidth]{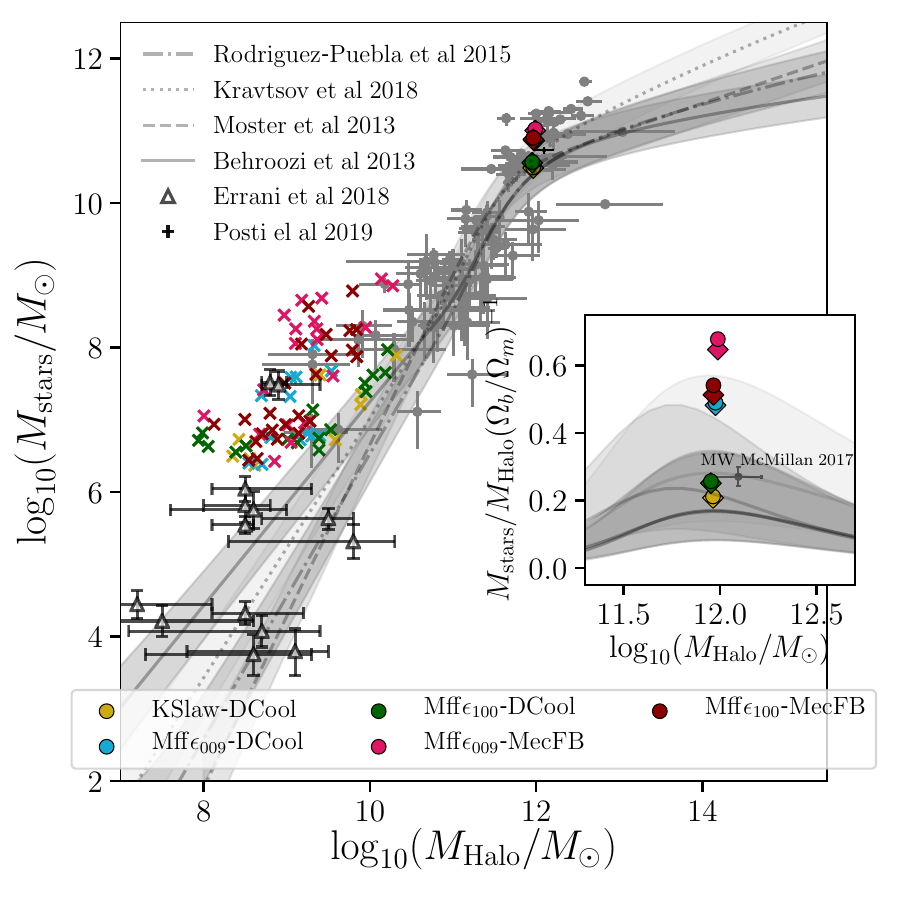}
    \caption{\label{fig:SHMR}}
  \end{subfigure}
  \label{fig:coffede}
    \begin{subfigure}[b]{0.45\textwidth}
       \includegraphics[width=\textwidth]{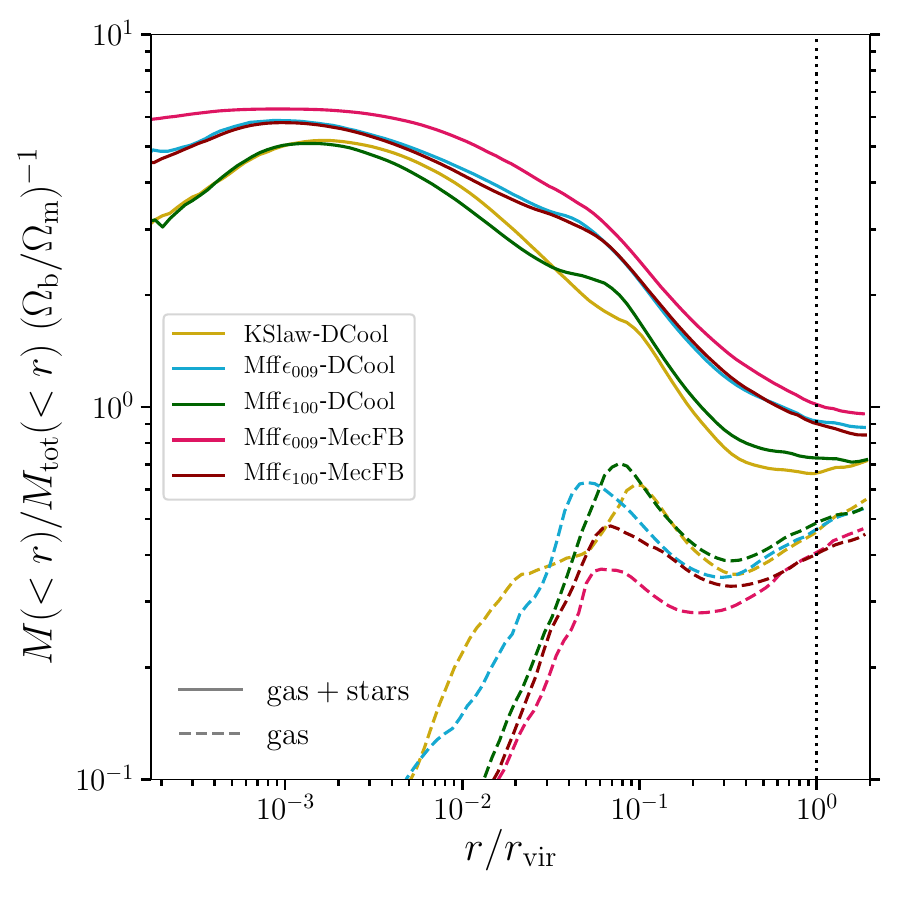}
        \caption{\label{fig:barfrac}}
    \end{subfigure}
  \caption{ (a) Stellar to Halo mass ratio at redshift 0 of the five runs. The stellar mass inside $r_{\mathrm{vir}}$ (circles) or inside $0.2\;r_{\mathrm{vir}}$(diamonds). The ratios for the satellite galaxies are shown in x's with colours corresponding to each run. Abundance matching relations are shown in grey \citep{behroozi2010comprehensive, munshi2013reproducing, rodriguez2015stellar, kravtsov2018stellar}. Observed stellar masses for nearby star-forming galaxies \citep{2019A&A...626A..56P} and Milky Way satellites \citep{2018MNRAS.481.5073E} with their estimated DM mass are shown in grey and black errorbars respectively. The reconstructed mass for the Milky Way \citep{Mcmillan2016} is also shown. (b) The baryonic fraction of the total contained mass (stars+gas+DM) in terms of the cosmological baryonic to total matter fraction with respect to the distance to the centre of the galaxy (solid lines). The gas fraction is also shown with respect to the total contained mass (dashed lines).  }

\end{figure*}

Additionally, we take $t_{\mathrm{sne}}=5$~Myr as an effort to suppress early star formation in the runs where mechanical feedback is used.\break

\section{Results}\label{sec:results}
  
We ran five simulations of the same galaxy, labelled \emph{Mochima}, a disc galaxy hosted by a Milky Way size halo inside a cosmological box of $\sim$ 36 Mpc of side, changing the star formation and the SN feedback recipes. The five runs have the following setup:
\begin{itemize}
\item \textbf{KSlaw-DCool}: Schmidt law and delayed cooling 
\item \textbf{Mff$\epsilon_{009}$-DCool}: multi-ff SF and delayed cooling with strong PSFB $\epsilon=0.09$
\item \textbf{Mff$\epsilon_{100}$-DCool}: multi-ff SF and delayed cooling with weak PSFB $\epsilon=1$
\item \textbf{Mff$\epsilon_{009}$-MecFB}: multi-ff SF and mechanical feedback with strong PSFB $\epsilon=0.09$

\item \textbf{Mff$\epsilon_{100}$-MecFB}: multi-ff SF and mechanical feedback with weak PSFB $\epsilon=1$

\end{itemize}

\subsection{Global properties}\label{subsec:Globalprop}

\subsubsection{Galaxy morphology}\label{sec:galmorpho}
To illustrate how the galaxy morphology is affected by the choice of sub-grid modelling of baryonic physics, in figure \ref{fig:maps}, we show maps, face on and edge-on, of the different runs at $z=0$. The upper row shows true color luminosity maps including dust obscuration in the SDSS bands created with SKIRT \citep{2015A&C....12...33B}, for these images a dust-to-gas ratio of 0.01 have been assumed in agreement with observation of local galaxies \citep{2013ApJ...777....5S} (for details on the production of these images see \citet{2017MNRAS.470..771T}). The bright blue regions denote recent star formations sites while the dark patches show the effect of the absortion by dust. The edge on view is naturally more affected by the dust absortion than the face on view, which is consistent with observations in the local universe where the redest galaxies obseved are typically edge on discs \citep{2013MNRAS.434.2503S}. Additionally in the middle and lower rows we show the respective gas density and temperature maps for all the runs.
As can be seen in figure \ref{fig:maps}, the Schmidt law star formation with the delayed cooling feedback (KSlaw-DCool) results in a well extended and diffused gas disc \footnote{Note that for this run the images have 60 kpc of side while for the other runs the side is of 50 kpc.}. From these maps, it seems that a density-driven star formation is not efficient enough to generate a stellar population in the outskirts of the disc where the gas is almost completely disrupted. The resulting stellar disc is very smooth, most of its star formation is concentrated in the bulge except for some punctual, faint, star formation sites in the outskirts of the disc. From $r=10$ kpc inwards, the gas disc is very thin, but the feedback is strong enough to push some of the gas outwards perpendicularly to the disc plane. 
The second and third columns of panels in figure \ref{fig:maps} show the same galaxy simulated with the multi-ff star formation and delayed cooling feedback variating the $\epsilon$ parameter. In the second column (Mff$\epsilon_{009}$-DCool) a strong PSFB is considered i.e. $\epsilon=0.09$ and the third column (Mff$\epsilon_{100}$-DCool) we use $\epsilon=1$ corresponding to a weak PSFB. For the strong PSFB case, the galaxy becomes less extended, and the gas ends up less diffused than in the fixed $\epsilon_{\mathrm{ff}}$ run (KSlaw-DCool). In this case, the spiral arms are denser in gas and more visibly populated by stars. Once the PSFB is factored out by setting $\epsilon=100\%$, we observe the extension of the gas disc, and a fainter star population. Additionally, fewer bright star formation sites are observed in the disc. After changing to the multi-ff star formation strategy the galactic stellar distribution becomes slightly clumpy and presents dense clouds of star-forming gas all over the disc. Many small and bright, young star regions can be seen all along the spiral arms, more so for the strong PSFB scenarios than for the weak scenarios. A massive bulge is observed in every run but with different temperature and mass distributions.

In this two runs with multi-ff SF and Delayed Cooling FB (Mff$\epsilon_{009}$-DCool and Mff$\epsilon_{100}$-DCool), the competition between star formation and feedback results in a thicker gas disc due to an evenly populated disc in stars that results in an evenly spread SN distribution. This can be seen in the temperature map and compared to the KSlaw-run where the temperature distribution is smoother and concentrated towards the centre.
  
  In the third groups of runs, we change the feedback strategy from Delayed Cooling to the mechanical feedback, and consider the strong and weak scenarios for the PSFB, Mff$\epsilon_{009}$-MecFB and Mff$\epsilon_{100}$-MecFB respectively. Judging qualitatively from the density maps, Delayed Cooling is more efficient at blowing out the gas vertically from the disc than the mechanical FB. 
  
  Having a strong PSFB with mechanical feedback (Mff$\epsilon_{009}$-MecFB) yields over-dense gas regions that are extremely efficient at forming stars. This means that a lower SF efficiency results in a weaker local FB unable to disrupt dense clouds, and this dense clouds become ultra-efficient SF regions. Such regions are seen as very bright spots in the luminosity map in figure \ref{fig:maps} for the Mff$\epsilon_{009}$-MecFB run. Consequently, the bulge in this run is the heaviest in stars, with respect to the other runs. On the other hand, we observe a drastically different situation when the weak PSFB scenario is considered (Mff$\epsilon_{100}$-MecFB). This is not surprising since, typically, higher values for $\epsilon$ are suggested in the literature \cite{2012ApJ...761..156F}. In this scenario, the disc is more extended, and no bright spots are seen in the luminosity map. However, the stellar distribution is very smoothly distributed while in reality, stars are seen to have a clumpy distribution.
   Generally, at $z=0$, the multi-ff star formation forms denser and well defined spiral arms that extend to the outskirts of the disc, contrary to what is observed in the Schmidt law.

\subsubsection{Stellar and gas mass fraction}
Abundance matching techniques between big volume cosmological simulations and galaxy surveys give an insight into the correspondence of halo mass to galaxy mass \citep{behroozi2010comprehensive, munshi2013reproducing, rodriguez2015stellar, kravtsov2018stellar}. However there are uncertainties within abundance matching techniques, coming either from the galaxy survey on the definition of the stellar mass and from counting issues inside the surveys, and on the simulation side from the cosmological parameters and the (not well understood) impact of baryonic physics in the halo properties. Therefore it is difficult to say whether comparing zoomed hydrodynamical simulations to stellar to halo mass ratio (SHMR) is a definite test of the reality of the results.

In figure \ref{fig:SHMR}, we show the relation between the stellar mass and the halo mass of the Mochima galaxy in our different runs. We show for each run two different definitions of the stellar mass, the full stellar mass inside $r_{\mathrm{vir}}$ (circles) and the stellar mass inside 20$\%$ of  $r_{\mathrm{vir}}$ (diamonds). Here we compare their SHMR with different semi-analytic abundance matching techniques \citep{behroozi2010comprehensive, munshi2013reproducing, rodriguez2015stellar, kravtsov2018stellar} and a set of carefully studied nearby star-forming galaxies \citep{2019A&A...626A..56P} shown as grey bands or grey points respectively.
We observe a good agreement between the SHMR and our simulations. Even if the mechanical feedback run with strong PSFB (Mff$\epsilon_{009}$-MecFB) ends up above the abundance matching prediction, it is perfectly consistent with the scatter in the observed galaxies. In figure \ref{fig:SHMR} we also show a frame that focuses on the region surrounding the central galaxy mass, here we show the resulting SHMR from a Milky Way mass model meant to fit constraints from photometric and kinematic observations \cite{Mcmillan2016} and how it compares to the Mochima galaxy different runs.

The run with delayed cooling FB have around $\sim 2\%$ of the total stellar mass is in satellite galaxies, thanks to the SN feedback efficiency in quenching the star formation in such galaxies. For the mechanical feedback runs, $\sim 5\%$ of the total stellar mass is inside satellites hinting that this feedback is not efficient enough to control the star formation in substructures. Too many satellites can form stars in this runs (Mff$\epsilon_{009}$-MecFB and Mff$\epsilon_{100}$-MecFB) compared to the threes runs with delayed cooling, this is evidenced in the difference of the resulting stellar mass inside $r_{\mathrm{vir}}$ with respect to the very inner stellar mass. To extend this argument, we use a the ROCKSTAR phase space temporal halo finder \cite{2013ApJ...762..109B} to find the DM substructures and select the subhaloes with a stellar counterpart and show their SHMR as x's in figure \ref{fig:SHMR}. We compare the found satellites in the simulations with the SHMR between the observed stellar masses for the satellites in the MW with their estimated DM mass using dynamical constraints and assuming cuspy profiles \citep{2018MNRAS.481.5073E}. We note that most of the low mass satellites observed in the MW fall close to our resolution limit (see section \ref{sec:thesimulation}). In particular two MW satellites, the Sagittarius dSph and Fornax (the two-point with the most massive stellar component)  exhibit an SHMR that is comparable to the satellites observed in all our runs even if far for the mean of the abundance matching prediction. However, we observe a systematically higher stellar mass in the detected satellites when compared to the abundance matching predictions. Taking into account the different sources of uncertainties, we consider that our satellites are in the ballpark of observations.
In figure \ref{fig:barfrac}, we show baryonic mass (stars+gas) fraction of the total mass of the halo (stars+gas+DM) as we increase the distance from the centre. As a check, it is shown in terms of the cosmological baryonic matter fraction so at the edge of the halo it should be equal to unity if the galaxy does not expel a significant amount of gas from the halo. The results are in agreement with what is expected, except for the cases of KSlaw-DCool and Mff$\epsilon_{100}$-DCool that fall slightly short but not enough to be considered in flagrant disagreement with the cosmological baryonic ratio.

\subsubsection{The star formation history}\label{sec:SFR}
\begin{figure*}
\centering 
\includegraphics[width=.98\textwidth]{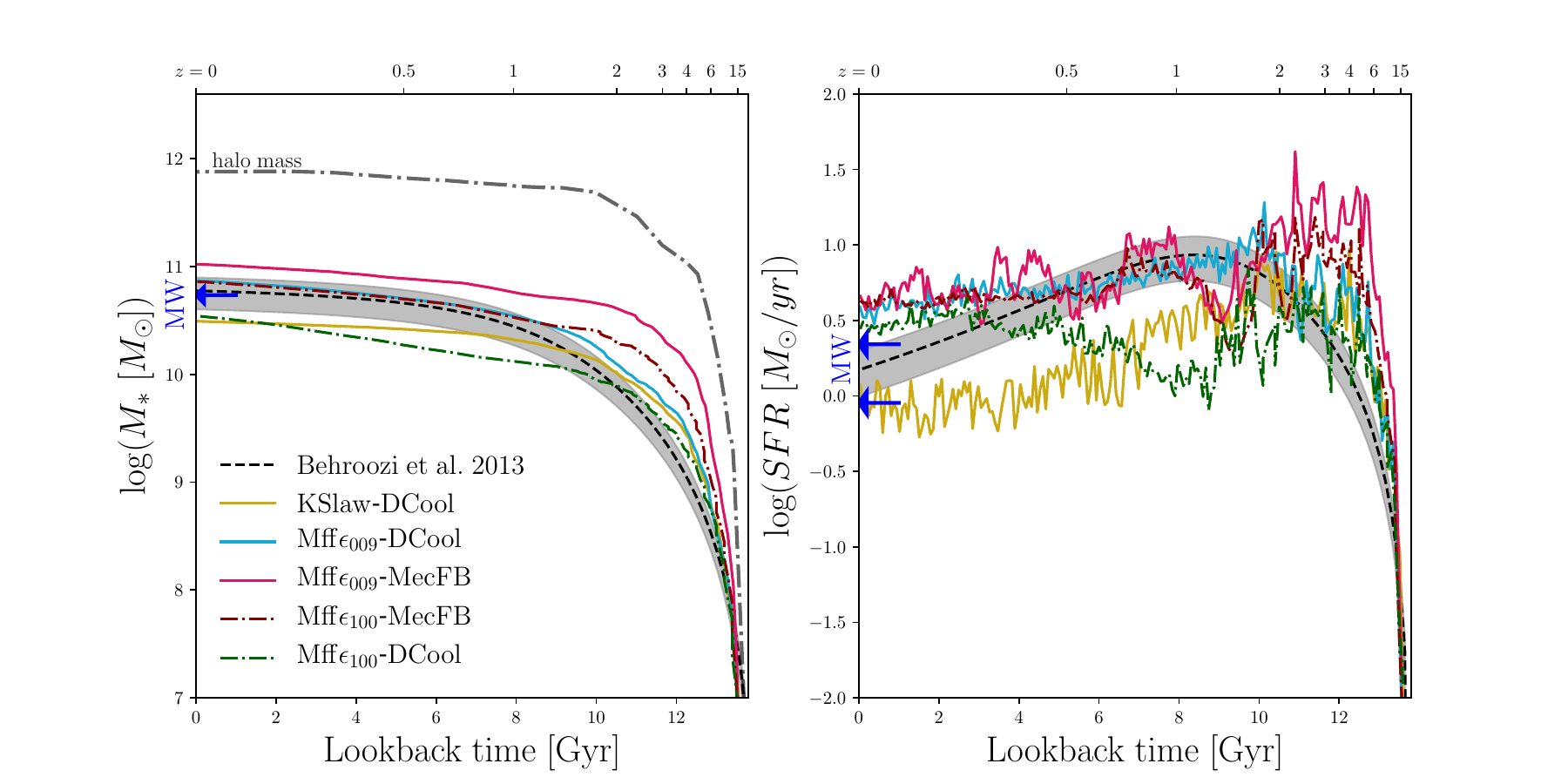}
	\caption{\label{fig:SFRH} Comparison of the galaxy stellar mass evolution (left) and star formation history (right) of the Mochima galaxy in lookback time in all the five runs. The stellar mass and ages are calculated for all the star particles inside the virial radius at $z=0$. Predictions from \citep{2013ApJ...770...57B} are shown in black dashed line with 1 $\sigma$ band, additionally, observations for the limits of SFR. Additionally, the evolution of the halo mass is shown (grey dot-dashed).The corresponding values for the MW today are shown by blue arrows from \citet{Mcmillan2016} for the stellar mass and from \citet{2010ApJ...709..424M} for the local SFR.  }

\end{figure*}

We also study the evolution of the stellar mass and the star formation rate (SFR) history of the simulated galaxies. For comparison, we use predictions for MW-like halos from semi-analytical models combining stellar mass function and halo merger histories \citep{2013ApJ...770...57B}, as shown in figure \ref{fig:SFRH}. It can be seen that for the run with constant $\epsilon_{\mathrm{ff}}$ over time and delayed cooling feedback (KSlaw-DCool), the SFR falls below the predictions after reaching a peak around $z\gtrsim 2$ at 10  M$_{\odot}$ yr$^{-1}$, while the peak value is that favoured by the model it is reached earlier. After the peak has been reached the SFR is quenched to one order of magnitude below what is predicted. One approach to deal with this discrepancies could be to calibrate the $\epsilon_{\mathrm{ff}}$ as it is the free parameter of this star formation strategy, on the other hand, an enhanced star formation efficiency might boost the rate of SN events, therefore, quenching the SFR. Alternatively, by switching to a variable $\epsilon_{\mathrm{ff}}$ in the multi-ff runs with delayed cooling, we observe a better agreement for the SFR history at $0.5<z<4$. Although an excess in the SFR is still observed for both recent and old stars in the system, the population of poorly regulated old stars, formed for $z>2$, will end up populating the stellar bulge, hence the mechanical feedback run with strong PSFB (Mff$\epsilon_{009}$-MecFB) where the highest number of old stars is seen, results in the most massive stellar bulge. While the mechanical feedback is able to regulate star formation in the last Gyrs, in good agreement with the equivalent run with delayed cooling feedback, it is not able to regulate the formation of early stars. This situation ends up assembling a massive galaxy that forms most of its stellar mass before $z=2$ following the growth of the dark matter halo, as shown in the left panel of figure \ref{fig:SFRH}. Here, it can be seen that the stellar mass of the mechanical feedback run rises very quickly before  $z=2$ where it slows down and remains almost constant, as opposed to the other three delayed cooling runs, where the steady growth of the stellar mass is observed $z<1$. The reduction of the PSFB for the mechanical FB run (Mff$\epsilon_{100}$-MecFB) reduces the early star formation but not enough to avoid the bulge; however, a significant reduction of the SFR is seen for $1<z<2$. In the case of the delayed cooling run with weak PSFB (Mff$\epsilon_{100}$-DCool), a significant reduction of SFR at all times is seen which results in the lightest galaxy with the multi-ff SF recipe.

If we consider the SFR today, by looking at the stars formed in the last 50 Myr, and compare it with the SFR today in the Milky Way (which is observed to be between 0.9 and 2.2 M$_{\odot}$ yr$^{-1}$ \citep{2010ApJ...709..424M} as shown in horizontal arrows in the right panel of figure \ref{fig:SFRH}), we see that today's rate in the Schmidt law SF run is of 0.5 M$_{\odot}$ yr$^{-1}$ in the galaxy. This is closer to the Milky Way's value than today's SFR in the four runs with the multi-ff star formation that have similar values of $\sim$4 M$_{\odot}$ yr$^{-1}$.

We find so far that the multi-ff star formation results in successful objects depending on the combination of the value of the PSFB and SN FB, explicitly for strong PSFB and delayed cooling and weak PSFB and mechanical FB, this situation highlights the high degeneracy and non-linearity of the galaxy evolution problem. However, we share the view exposed in \citet{Mollitor2014} and \citet{2018MNRAS.473.4077P} where it is argued that due to the various sources of uncertainties, this type of comparisons needs to be taken with caution.

\subsubsection{The Kennicutt-Schmidt relation}
\begin{figure}
\includegraphics[width=0.95\columnwidth]{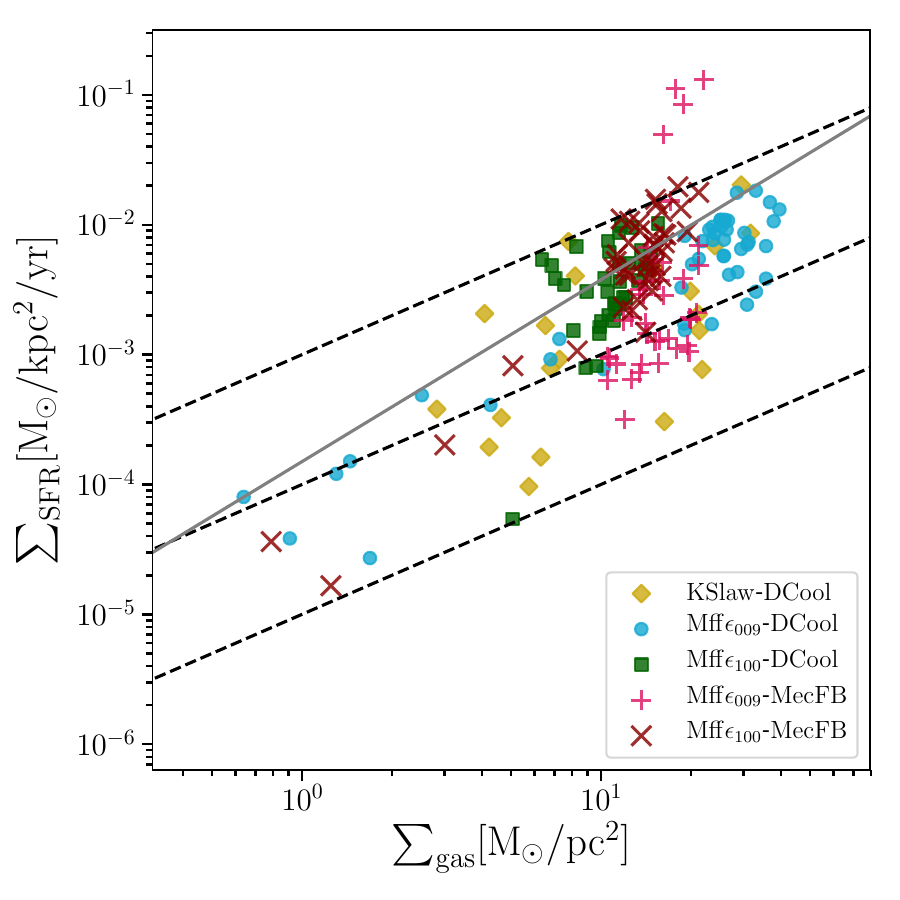}
\caption{The Kennicutt-Schmidt relation for different baryonic physics combinations for the last 50 Myr. The solid line shows the empirical Kennicutt law \citep{kennicutt1998global}, the dashed lines show the 0.1, 1 and 10 $\%$ efficiency of star-formation. }\label{fig:KSL}
\label{fig:example}
\end{figure}

The SFR surface density, $\sum_{\mathrm{SFR}}$, at large scales in the local universe is observed to follow the global Kennicutt-Schmidt (KS) relation for star-forming galaxies \citep{kennicutt1998global}. Where the SFR surface density scales as a power law of the gas surface density, $\sum_{\mathrm{SFR}} \propto \sum_{\mathrm{gas}}^{1.4}$. In figure \ref{fig:KSL}, we show the relation between the gas and the SFR surface density for different strategies of baryonic physics in the same galaxy in the last 50 Myr. We show the empirical Kennicutt-Schmidt relation normalized to the Chabrier IMF (see \cite{dalla2012simulating})  in a solid line and the star formation efficiency required to consume 10, 1 and 0.1$\%$ of the gas in dashed lines. The gas and SFR surface densities are averaged over tori in the galactic plane centred in the galactic centre with equal azimuthal bins of $\Delta r = 500$ pc and a 2 pc heights. 

We observe rough agreement of all our baryonic physics strategies with observation at surface densities of $\sum_{\mathrm{gas}}     \approx$ 20 M$_{\odot}$ pc$^{-2}$. The similarity of the Schmidt law SF strategy run (KSlaw-DCool) and the KS relation is somewhat expected due to the dependence of the SFR to the gas density, $\dot{\rho_{*}}\propto \rho_{\mathrm{gas}}^{1.5}$ (see equation \ref{eq:starformation}) when $\epsilon_{\mathrm{ff}}$ is kept constant, contrary to the case of the multi freefall star formation where this is no longer true, and the dependence is more complex. 

Comparing the four cases where the multi-ff SF is used, we can observe the effect of the different feedback implementations. The two degenerated successful runs (Mff$\epsilon_{009}$-DCool and Mff$\epsilon_{100}$-MecFB) reproduce well the slope of the KS relation but with slightly lower efficiency. While the mechanical feedback with strong PSFB (Mff$\epsilon_{009}$-MecFB) is only allowing star formation in high gas surface density regions, this induces very efficient gas consumption in the central regions of the galaxy and a very massive stellar bulge. Similarly, the delayed cooling run with weak PSFB (Mff$\epsilon_{100}$-DCool) allows star formation in very dense regions but without the over-efficient clouds of the Mff$\epsilon_{009}$-MecFB run.

\begin{figure*}
  \centering
  \begin{subfigure}[b]{0.46\linewidth}
    \includegraphics[width=\linewidth]{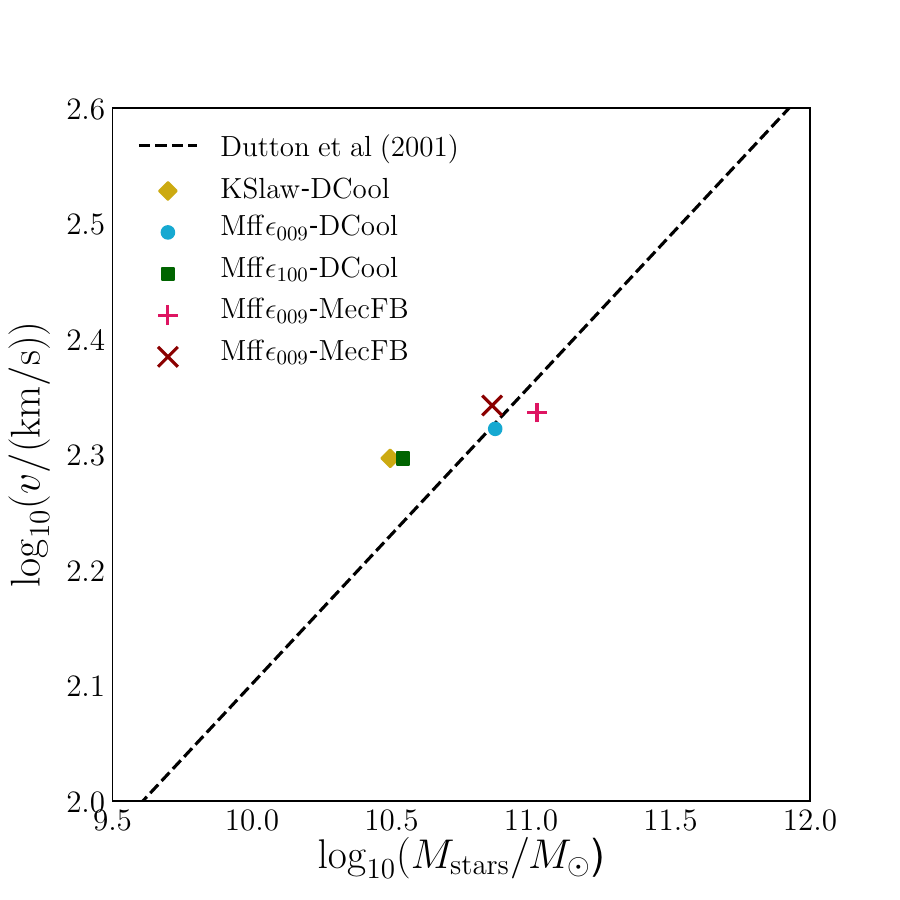}
    \caption{}
  \end{subfigure}
  \label{fig:coffeef}
    \begin{subfigure}[b]{0.46\linewidth}
       \includegraphics[width=\linewidth]{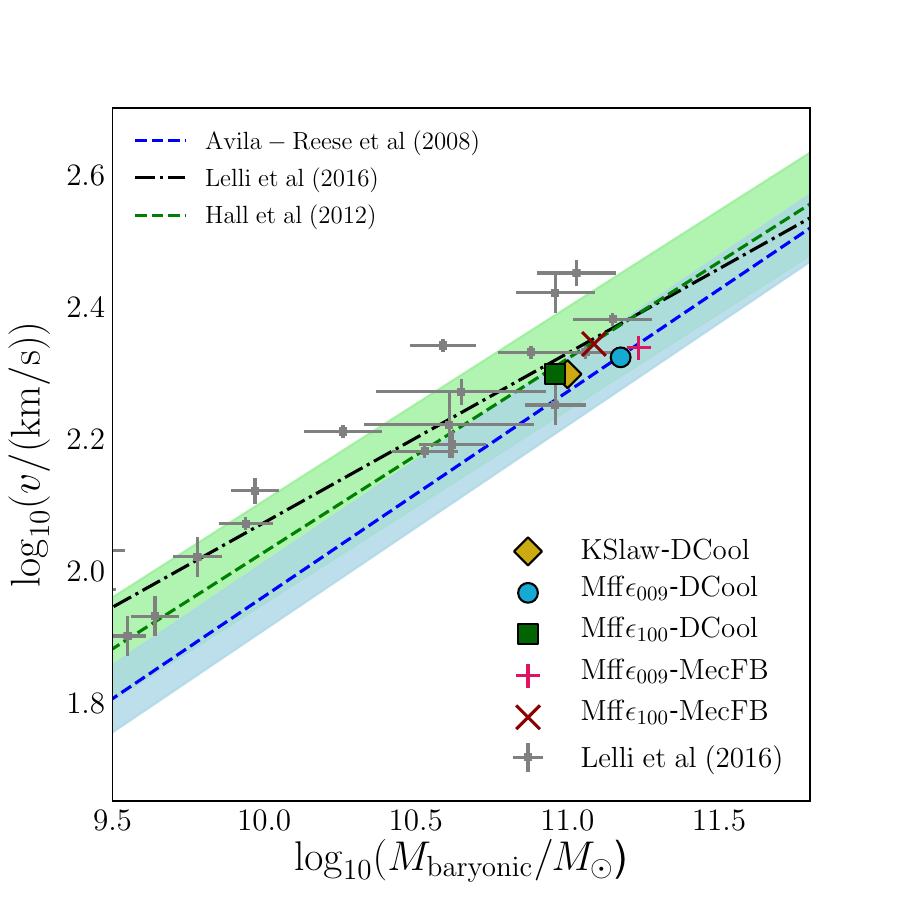}
        \caption{}
    \end{subfigure}
  
  \caption{(a) Tully-Fisher relation where the stellar mass computed inside 10$\%$ of the $r_{\mathrm{vir}}$ and the rotation speed $v$ correspond to the maximal tangential stellar speed observed in the disc. A fit on observation from  \citet{2011MNRAS.410.1660D} is shown in the dashed line. (b) The baryonic Tully-Fisher relation is analogous to the computation shown in (a) but now including the gas. The dashed lines show fits observations by \citet{2012MNRAS.425.2741H,2008AJ....136.1340A,2016AJ....152..157L} additionally the observations of the SPARC survey are shown in the grey errorbars \citep{2016AJ....152..157L}. }
  \label{fig:TF}
\end{figure*}

\subsubsection{Tully-Fisher relation}

\label{subsubsec:dynandobs}

In figure \ref{fig:TF}, we show the Tully-Fisher (TF) relation in the left and the baryonic TF relation on the right at redshift 0. We used as the mass of the central galaxy the mass contained within 10$\%$ of the virial radius ($r_{\mathrm{vir}}$). The circular velocity is calculated from the contained mass as before for the radius in the disc where it reaches a maximum. We include several lines representing the best fit from observations in \cite{2011MNRAS.410.1660D} for the TF and in \cite{2008AJ....136.1340A,2012MNRAS.425.2741H} for the baryonic TF. The best agreement with these observations happen for the galaxy simulated with the multi-ff SF and the delayed cooling feedback and strong PSFB (Mff$\epsilon_{009}$-DCool), but all five galaxies are consistent with the dispersion of the observational points.

\begin{figure*}
  \centering
  \begin{subfigure}[b]{0.32\textwidth}
    \includegraphics[width=\textwidth]{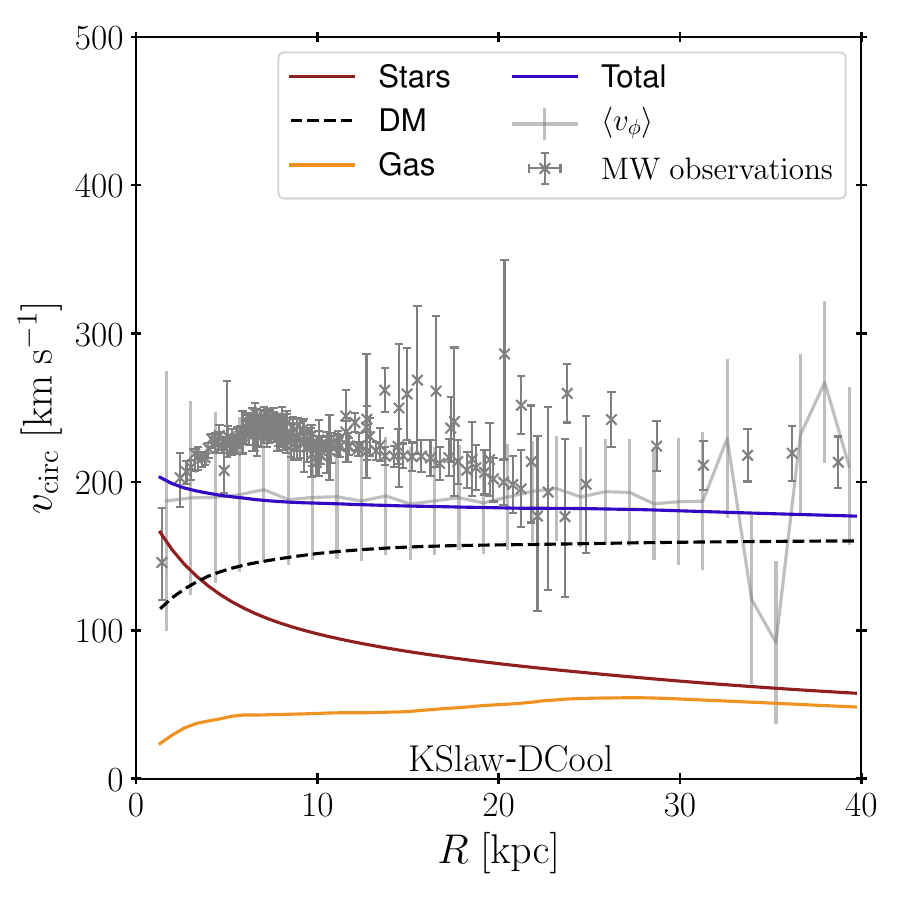}
  \end{subfigure}
  \label{fig:coffefe}
    \begin{subfigure}[b]{0.32\textwidth}
       \includegraphics[width=\textwidth]{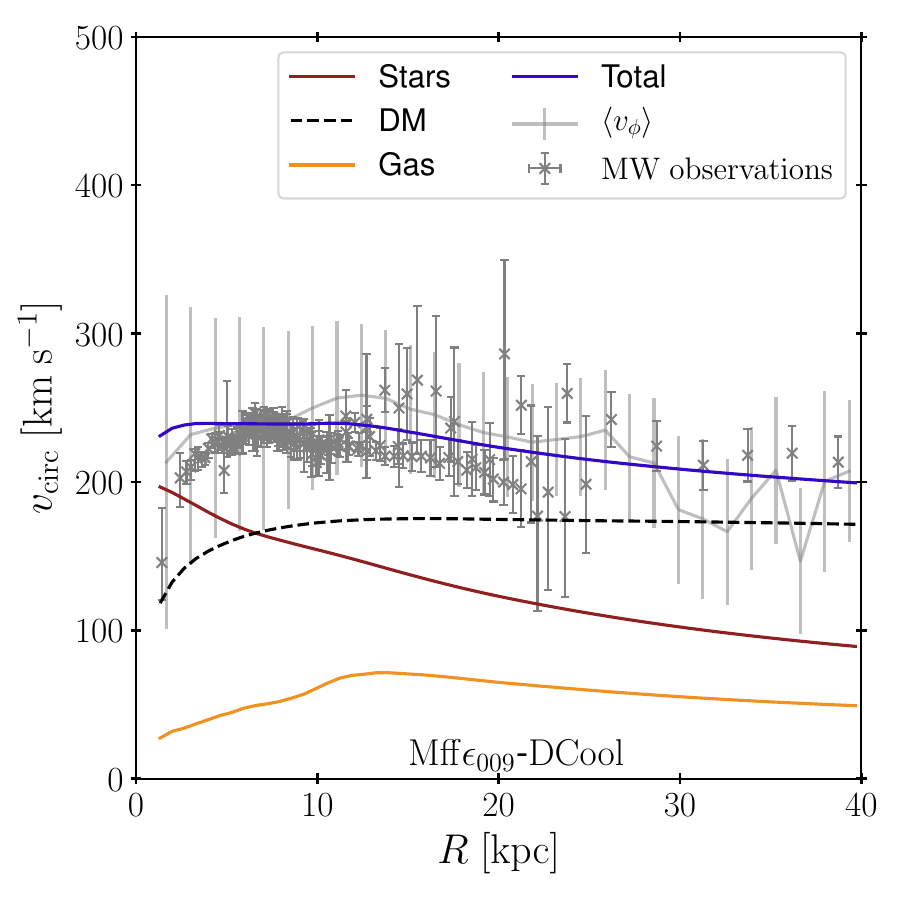}
    \end{subfigure}
  \begin{subfigure}[b]{0.32\textwidth}
    \includegraphics[width=\textwidth]{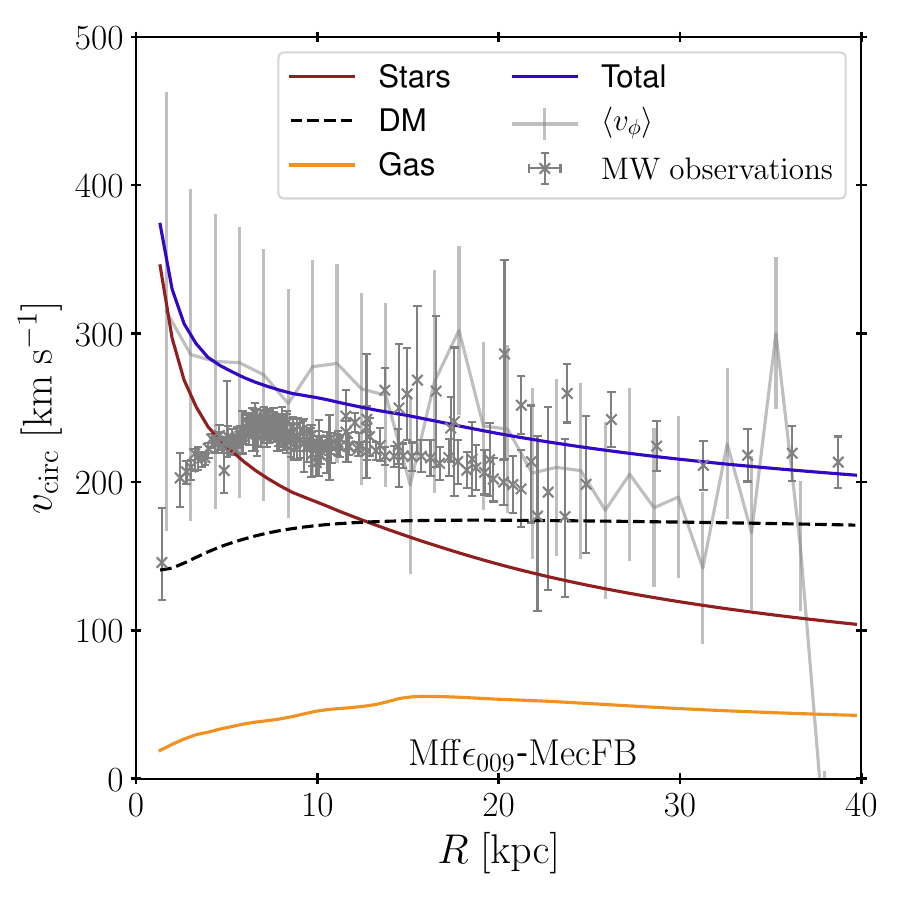}
  \end{subfigure}
    \begin{subfigure}[b]{0.32\textwidth}
    \includegraphics[width=\textwidth]{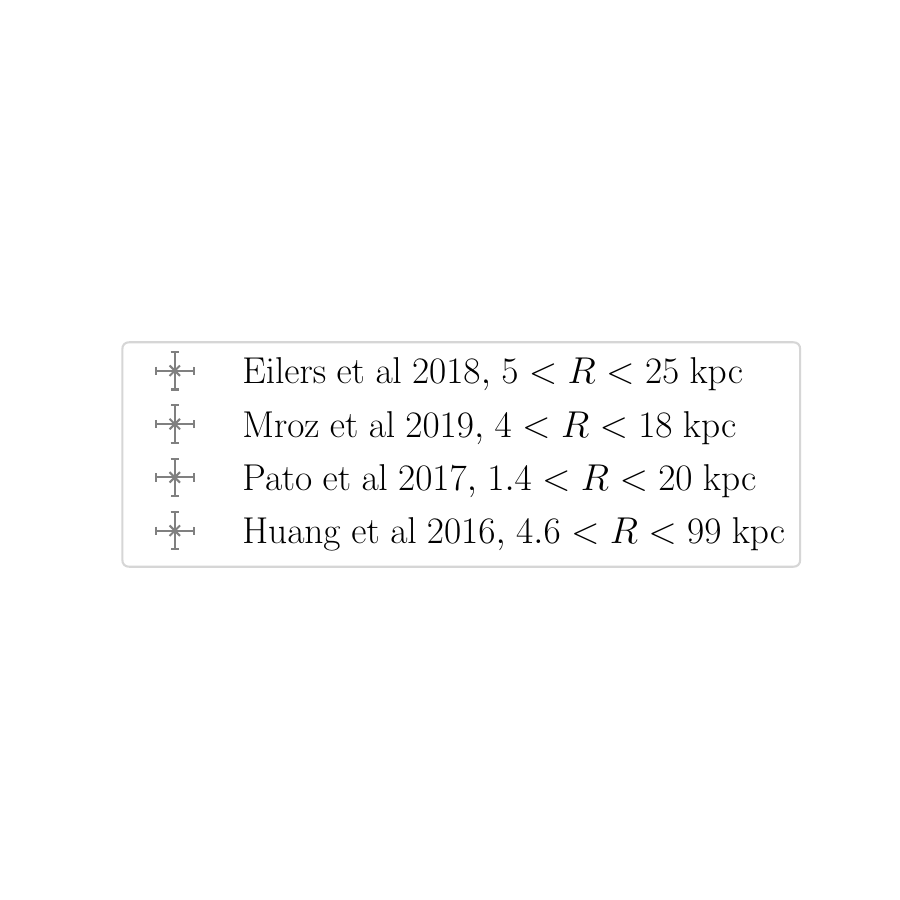}
  \end{subfigure}
  \begin{subfigure}[b]{0.32\textwidth}
    \includegraphics[width=\textwidth]{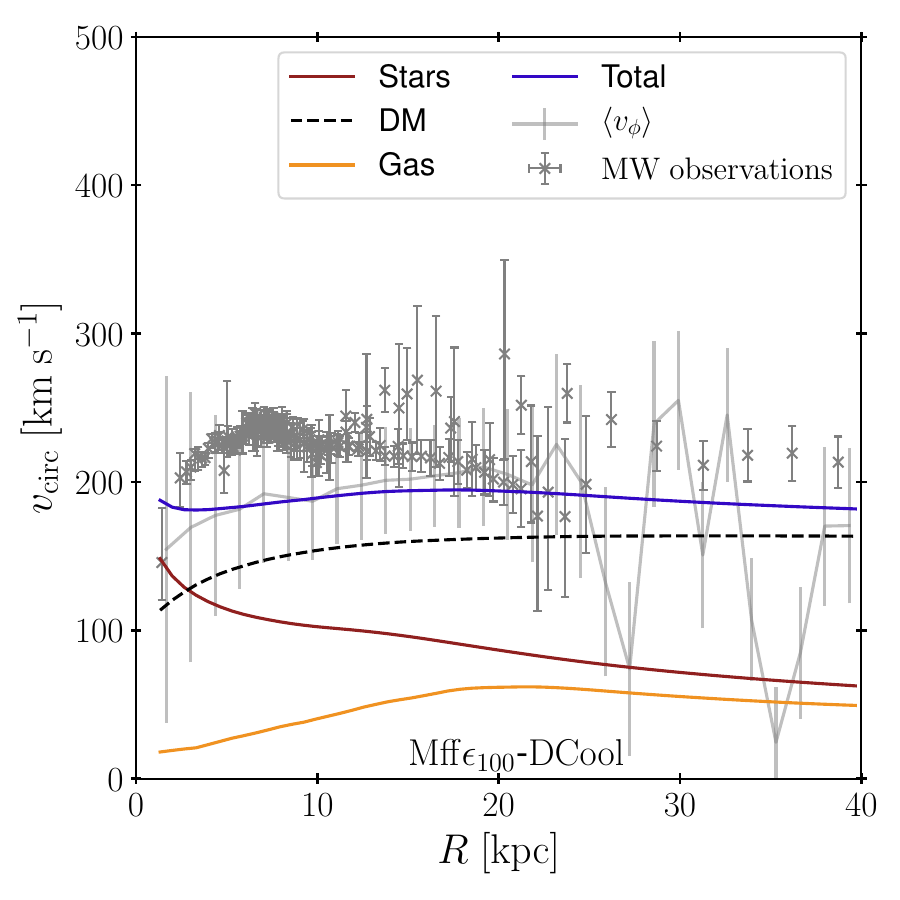}
  \end{subfigure}
    \begin{subfigure}[b]{0.32\textwidth}
    \includegraphics[width=\textwidth]{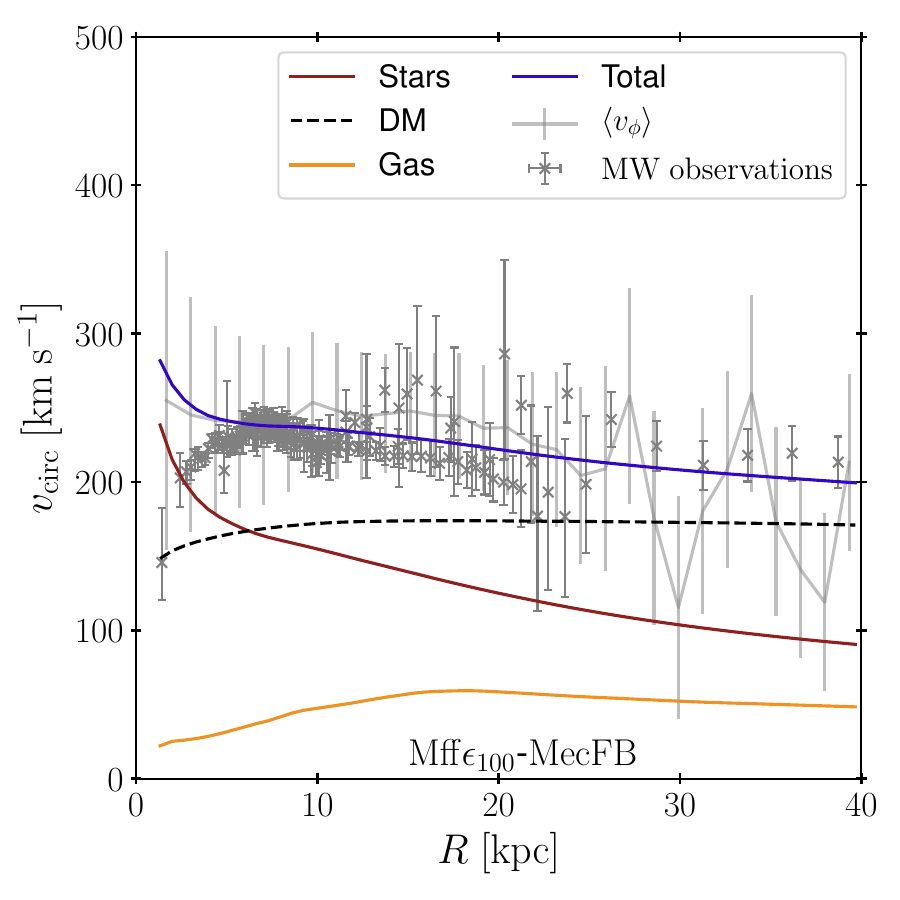}
  \end{subfigure}
  \caption{Rotation curves computed from the contained mass at a radius R of dark matter (black dashed), stars (red solid), gas (yellow solid) and the total mass (blue solid). Tangential velocities for the stars in the simulation is shown in error bars. Observations from the Milky Way are shown in grey errorbar x's form a compilation of MW observations  \citep{2016MNRAS.463.2623H,2017SoftX...6...54P,2019ApJ...871..120E,2019ApJ...870L..10M}. }
  \label{fig:RC}
\end{figure*}

\subsection{Local aspects}\label{subsec:Localprop}
\subsubsection{Rotation curves}
In figure \ref{fig:RC} we show the rotation curves (RC) of the five runs built either by the contained mass per radius for each and all the components ($v_c = \sqrt{G M(r)/r}$ where $M(r)$ is the contained mass inside $r$) or with the actual tangential velocities in the stars of the galaxy corrected for asymmetric drift correction \citep{Binney} in order to be consistent with observations. For comparison, we show a compilation of observations of the stars in the MW \citep{2016MNRAS.463.2623H,2017SoftX...6...54P,2019ApJ...871..120E,2019ApJ...870L..10M} in grey errorbar points. We see that for the case of the Schmidt law run (KSlaw-DCool) and the weak PSFB multi-ff run with delayed cooling (Mff$\epsilon_{100}$-DCool) the final galaxy is not massive enough to generate sufficient angular velocity in stars. The other three runs yield comparable tangential velocities of the stars to that of the MW disc around 10 kpc, but the presence of the massive bulge generates a violent rise in the RC that does not agree with MW observations. In the particular case of two runs with mechanical feedback (Mff$\epsilon_{009}$-MecFB and Mff$\epsilon_{100}$-MecFB), the bulge is so massive that a spike is observed towards the centre of the galaxy in the RC. Here, a particular difference is seen between the two successful runs, Mff$\epsilon_{009}$-DCool and Mff$\epsilon_{100}$-MecFB), in the central region where the latter shows an asymptotic spike in the centre, nevertheless both galaxies show impressive agreement with MW observation for $R>5$ kpc. We focus on the MW for this comparison, nevertheless, it is worth remarking that rotation curves of other galaxies could exhibit a better agreement with our simulations.

\begin{figure*}
  \centering
  \begin{subfigure}[b]{0.46\linewidth}
    \includegraphics[width=\linewidth]{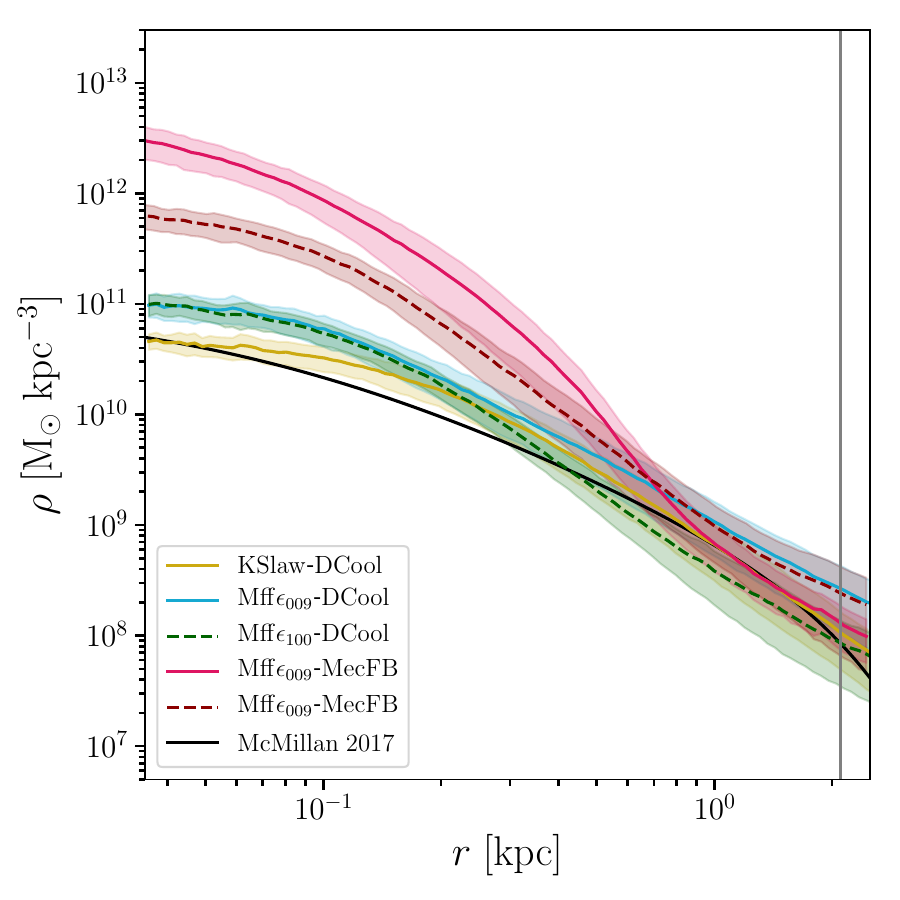}
    \caption{}
  \end{subfigure}
  \label{fig:coffeeg}
    \begin{subfigure}[b]{0.46\linewidth}
       \includegraphics[width=\linewidth]{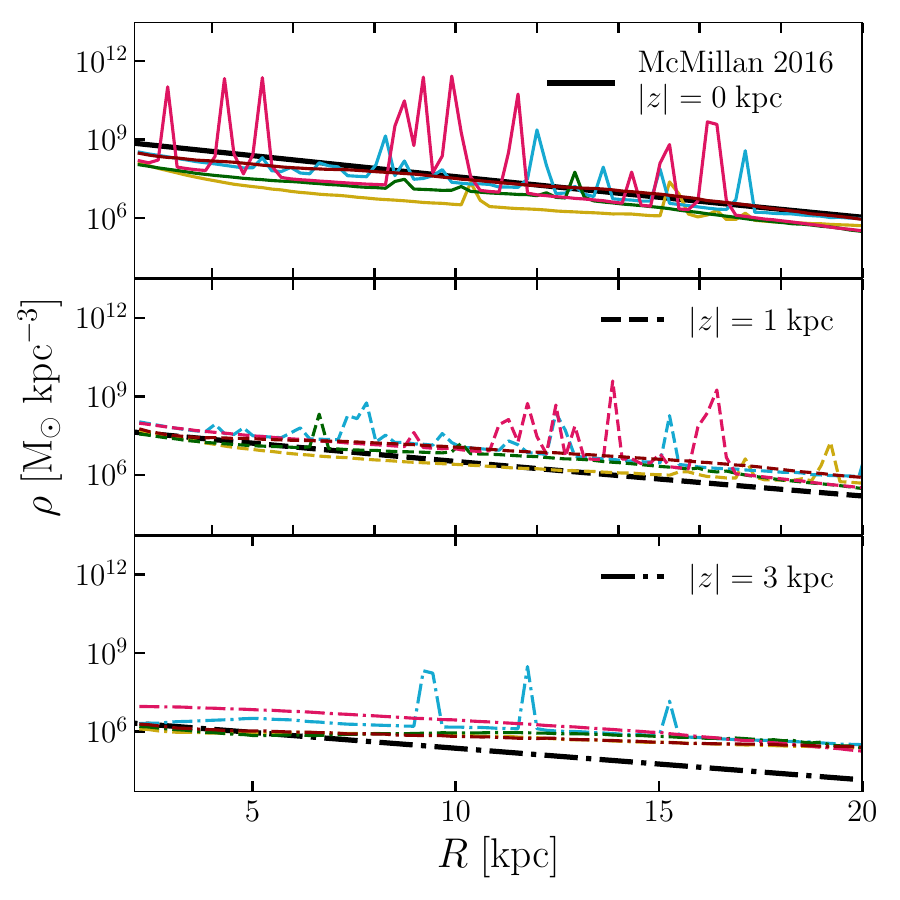}
        \caption{}
    \end{subfigure}
  
	\caption{\label{fig:bulgedisc} Density profile of the stellar distributions of the bulge (left), and disc (right) compared to the stellar mass model meant to fit observational constraints of the MW \citep{Mcmillan2016}. The comparison for the bulge is made in spherical coordinates, and for the disc in cylindrical coordinates fixing the value of z for the theoretical curve as denoted in the upper right corner of each panel, for the simulations, we use a bin of $\Delta z = 1$ kpc around the fixed value for the model.  }
 
\end{figure*}
\subsubsection{Stellar distribution}\label{sec:StellarDist}

The mass distribution of the Milky Way can be modelled to fit different observational photometric and kinematic constraints \citet{2011MNRAS.414.2446M,Mcmillan2016}. We compare our simulations to such models, and the results are shown in figure \ref{fig:bulgedisc} for the stellar bulge and disc, it is worth noting that there is no bar present in any of our runs. In the case of the bulge, shown in the left, we consider equation 1 of \cite{Mcmillan2016} on the spherical limit ($q=1$) and compare it to the spherically averaged stellar density. The relevant range in $r$ for this comparison spans from our resolution limit $\Delta x = 35 $ pc up to the $r_{\mathrm{cut}}= 2.1$ kpc (gray vertical line in figure \ref{fig:bulgedisc}). As mentioned above the obtained stellar population in the bulge exceeds that of what is expected for the Milky Way in all of our runs, in particular, the runs with mechanical FB presents a density profile around 10-20 times denser than what is predicted by the model at the resolution limit and with a similar ratio up to $r_{\mathrm{cut}}$. On the other hand, the three runs with Delayed Cooling present less departure from the model, and a better agreement is found in the Schmidt law SF run (KSlaw-DCool). In the case of the disc, we keep the full axisymmetric form of equation 3 in \citet{Mcmillan2016}. We add the thick and thin stellar discs of the model and show comparisons to the resulting stellar density in the disc with respect to the cylindric radius $R$ keeping $|z|$ constant at 0, 1 and 3 kpc.  In the case of the simulations, we use a bin in $|z|$ centred in the same values with 1 pc of width in $|z|$ and show the cylindrically averaged stellar density with respect to the $R$ in these bins in $z$. The results are shown in the right panel of figure \ref{fig:bulgedisc}. The best agreement for the stellar disc density at $|z|=0$ is obtained for the two successful runs (Mff$\epsilon_{009}$-DCool and Mff$\epsilon_{100}$-MecFB). In all runs, a thicker disc than the Milky Way disc is found. This is most likely due to a resolution effect, and even if we resolve the scale height of the thin and thick disc with ~8 and ~25 cells respectively, this might not be enough to resolve the full gas dynamics inside the galactic disc. In the next section, we focus on gas dynamics of the star-forming cells and compare with observations of regions of similar size either in the Milky way or nearby spiral galaxies.

\subsubsection{Star formation sites: gas features and observations}

Recent high-resolution observations of molecular clouds in the MW or in nearby galaxies together with the resolution achieved in the three Mochima simulations constitute an interesting framework to study the performance of our sub-grid physics implementations as compared to observed interstellar medium (ISM) physics. During the runs presented in this work, we have stored hydrodynamical quantities present in the gas cell, i.e. density, volume, temperature, and velocity dispersion at the moment where star formation is about to happen. In this section, we compare the hydrodynamical features of the star-forming cells obtained with the different sub-grid physics implementations. We also study how the gas of the star-forming cell compares to observations of star-forming regions in the MW or M51.
\begin{figure*}
\centering
  \begin{subfigure}[b]{0.32\textwidth}
    \includegraphics[width=\textwidth]{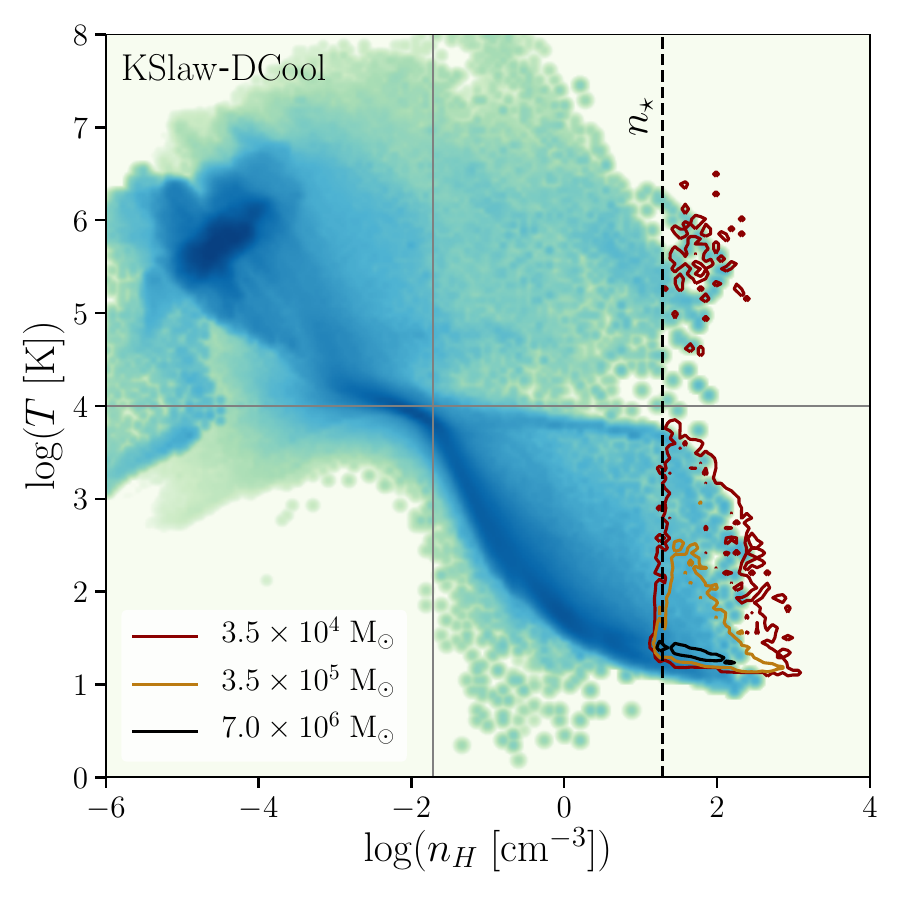}
  \end{subfigure}
  \label{fig:cogffee}
    \begin{subfigure}[b]{0.32\textwidth}
       \includegraphics[width=\textwidth]{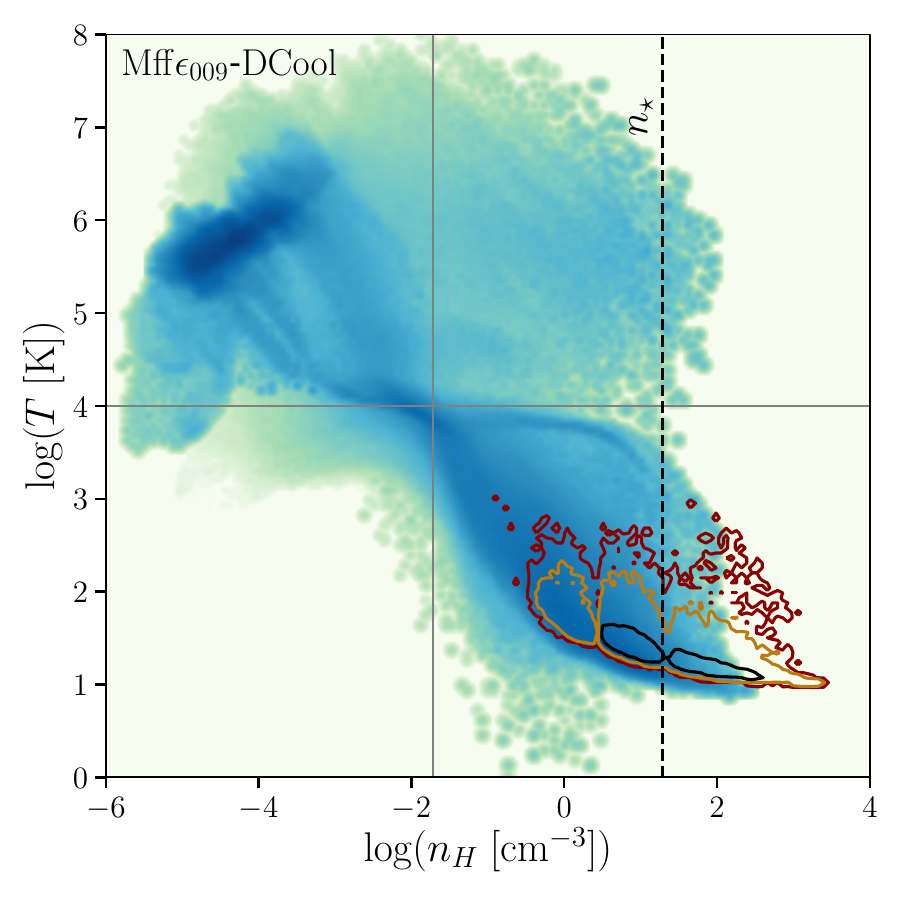}
    \end{subfigure}
  \begin{subfigure}[b]{0.32\textwidth}
    \includegraphics[width=\textwidth]{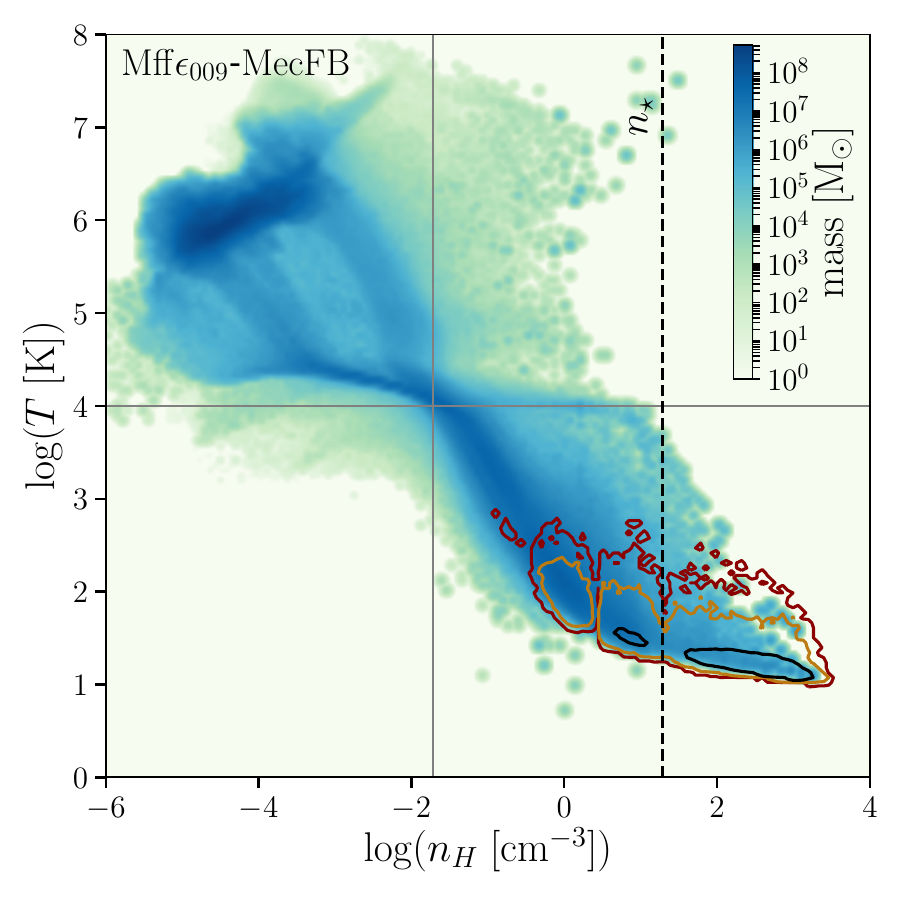}
  \end{subfigure}

  \label{fig:cogffee}
    \begin{subfigure}[b]{0.32\textwidth}
       \includegraphics[width=\textwidth]{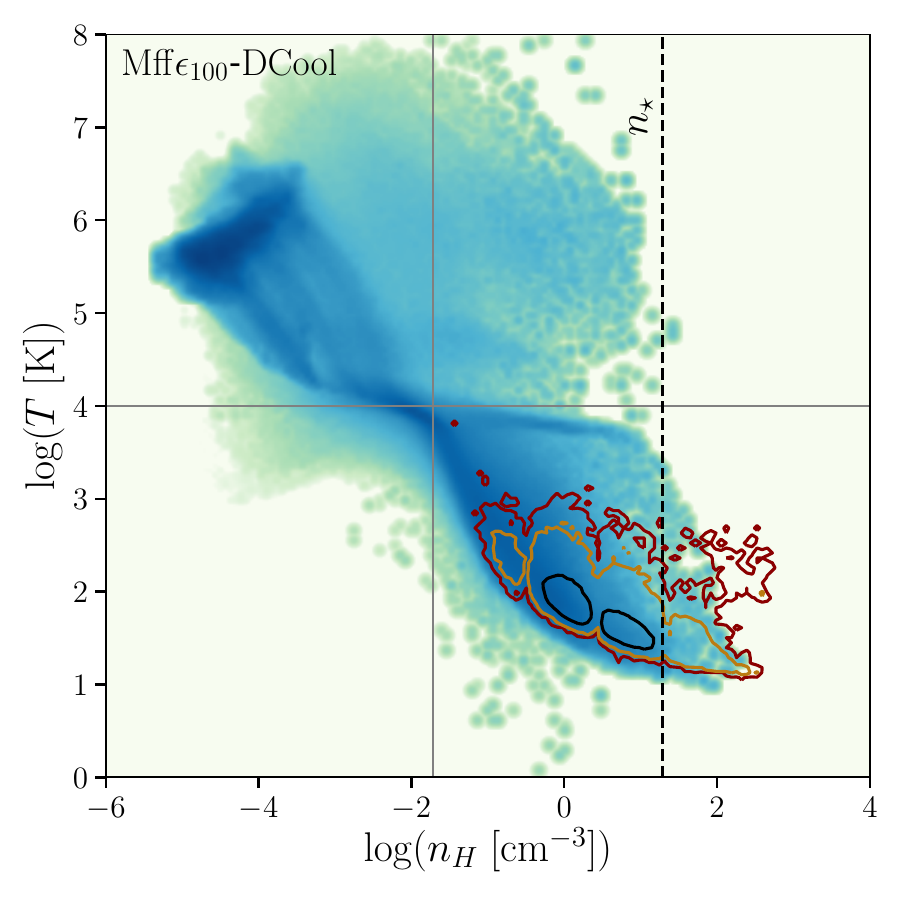}
    \end{subfigure}
  \begin{subfigure}[b]{0.32\textwidth}
    \includegraphics[width=\textwidth]{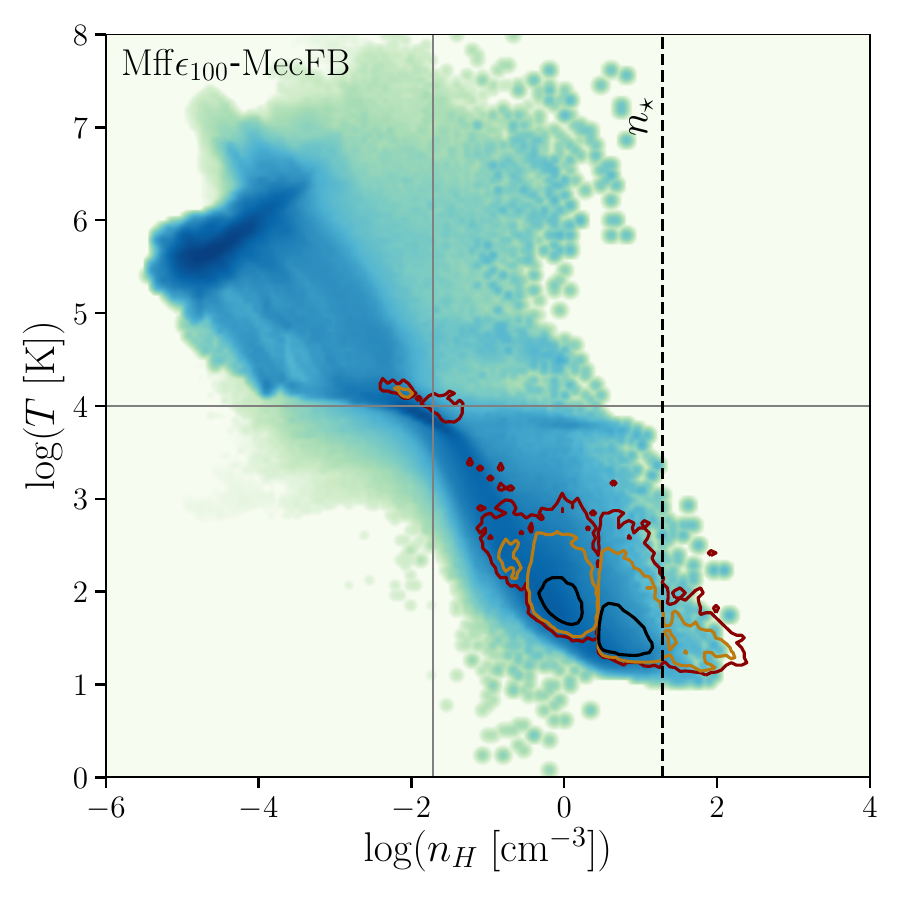}
  \end{subfigure}
\caption{\label{fig:PhaseD} 2D histogram of the density-temperature phase diagram of the gas inside $r_{\mathrm{vir}}$ at $z=0$ for the five runs. The density is shown as hydrogen density per cm$ ^3$. The contours indicate position in the digram of the star-forming cells for the last 500 Myr of different values of stacked stellar mass as shown in the legend.}
\end{figure*}

We start our study of the gas features by looking at its density temperature distribution, which is shown in figure \ref{fig:PhaseD} for all gas cells inside the virial radius of each galaxy. The density temperature diagram can be understood by following the treatment presented in \cite{2019MNRAS.485.2511T}, we further simplify this approach by dividing the diagram into four quadrants, hot and cold gas separated by $T=10^4$ K as discussed in \cite{2019MNRAS.485.2511T} and low density and high-density gas separated $0.01 \%$ of the density threshold imposed on the Schmidt law star formation (see section \ref{sec:SFR}). These two boundaries are shown as horizontal and vertical grey solid lines in the figure. Here, we can identify the gas belonging to the hot circum-galactic medium (CGM) as the low-density hot gas, this gas comes from the intergalactic medium (IGM) as low-density cold gas and after shock-heating becomes the hot CGM. The cold IGM gas could also be directly accreted as cold gas into the cold, dense quadrant and join the cold ISM. It is this cold, dense gas in the cold ISM that is available for star formation and is eventually reheated by the SN feedback. Gas in the bottom of the lower right quadrant when subject to SN feedback undergoes a temperature increase and turns into either very hot and dense clouds populating the hot ISM inside the disc, or into clouds that would reach temperatures of a few thousand Kelvin that subsequently expands reaching lower densities. Toward the crossing of the two boundaries, where they would move back into the star-forming gas as it cools down. In reality, when the gas comes into the star-forming region after being reheated by SN feedback, it will be metal-rich, from the SN explosion, and give rise to second-generation stars such as the Sun.

In figure \ref{fig:PhaseD} we also show the density temperature distribution of the star-forming cells in isocontours corresponding to regions in the diagram that have formed 1, 10 or 200 times the mass of the smallest star particle in the last 500 Myrs. We use the star-forming cells of the last 500 Myrs in each simulation to increase statistics, while the gas diagram corresponds to the galaxy at $z=0$. The contours are built with the gas cell features right before the gas is turned into stars. We observe that the Schmidt law star formation (KSlaw-DCool) generates stars in gas that belongs to the hot ISM and the feedback heated gas, i.e. gas that is too hot to be forming stars. Within this implementation, there is no regulation for temperature effects given that the only criterium to turn gas into stars is density. However, dense hot gas is not very likely to stay in this state for long due to radiative cooling. Therefore, very few cells will form stars in the hot ISM. The Schmidt-law star formation is forming stars within all the available gas above the density threshold regardless of its temperature, as shown in figure \ref{fig:PhaseD}. This issue is solved in the four runs with the multi-ff star formation, where no stars are formed in hot gas given that gas with high temperatures is turbulent and can support gravitational collapse. However in the run with mechanical FB and weak PSFB (Mff$\epsilon_{100}$-MecFB) some stars are formed with gas that lies in the intersection of all four quadrants, except for this case, since in the multi-ff model there is virtually no hard density threshold, a distribution of star-forming cells that is wider in density than in temperature is observed. Only cold gas is forming stars over two orders of magnitude in density\footnote{The multimodality of the distribution is related to resolution and the refinement strategy in RAMSES}.    

The hot ISM (upper right quadrant in figure \ref{fig:PhaseD}) has a higher population for the delayed cooling runs than in the mechanical feedback runs, and this supports our initial assessment that the former is more efficient at reheating the ISM than the latter. In particular, when combined with the multi-ff star formation, delayed cooling and strong PSFB (Mff$\epsilon_{009}$-DCool), the heating of the gas all along the disc is very efficient (see figure \ref{fig:maps}) resulting in a higher number of gas cells populating the hot ISM in the disc.

\begin{figure*}

  \centering
  \begin{subfigure}[b]{0.46\linewidth}
    \includegraphics[width=\linewidth]{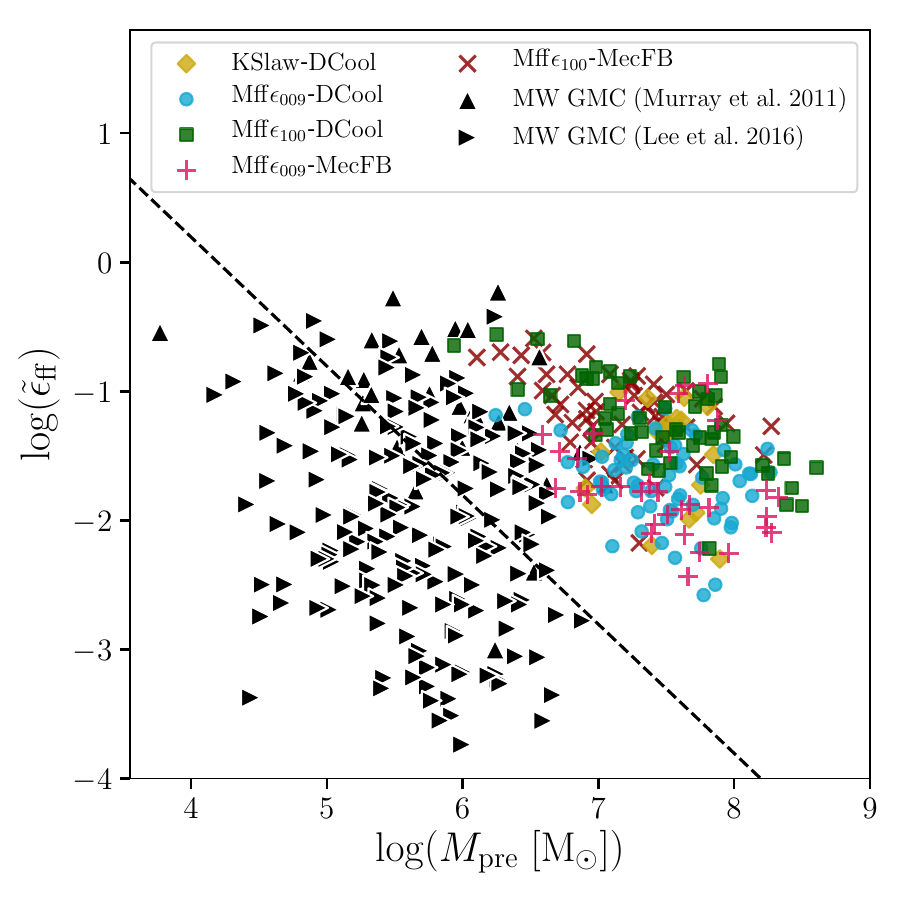}
    \caption{}
  \end{subfigure}
  \label{fig:coffaee}
    \begin{subfigure}[b]{0.46\linewidth}
       \includegraphics[width=\linewidth]{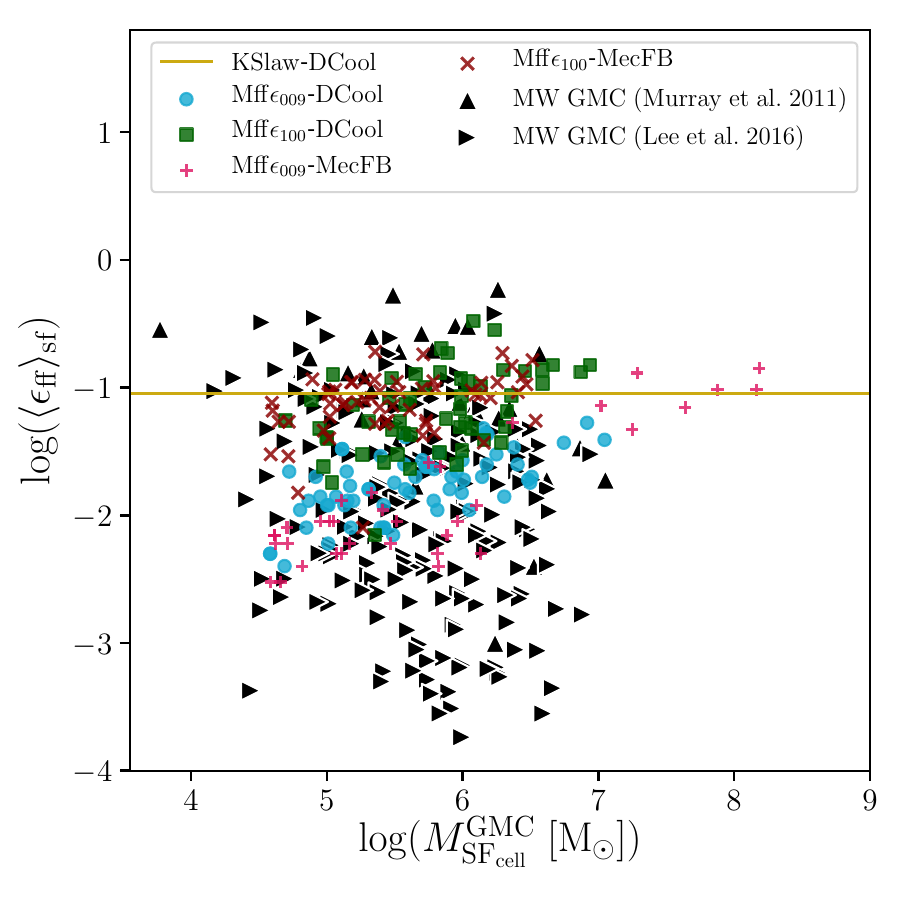}
        \caption{}
    \end{subfigure}
\caption{\label{fig:effvsMass} Two ways of comparing The star formations with MW observations from \citet{2011ApJ...729..133M,2016ApJ...833..229L}. In the Left, reconstructed efficiency following the observation approach. The observations for Milky Way giant molecular clouds are shown in the black triangles, in the right, the mean efficiency of the star-forming cells inside each GMC calculated by the multi-ff model with respect to the sum of their masses.  }
\end{figure*}

Observations of star formation regions in nearby spirals and MW clouds combine different wavelengths to relate the SFR and cloud mass to generate the star formation efficiency \citep{2019A&A...625A..19Q}. We use observations of the star formation efficiency with a resolution that ranges from 40 pc to 100 pc and compare them to the efficiency in the star-forming cells of the last 500 Myr in the five simulations. In figure \ref{fig:effvsMass}, we compare the SF efficiency as a function of the cloud mass previous to the birth of the star in the star-forming cells with observations from \cite{2011ApJ...729..133M,2016ApJ...833..229L}. In the left panel, we compare observations with the molecular clouds in the simulations. These molecular clouds are detected using the on-the-fly clump finding module PHEW \citep{2015ComAC...2....5B} inside RAMSES, similarly to the treatment in \citet{2019MNRAS.486.5482G}. The PHEW algorithm works by identifying AMR cells with densities above a predefined threshold, then, clumps are built by grouping together all nearby dense cells. Finally, clumps are merged if they are separated by a density saddle that is larger than a parameter $\rho_{saddle}$. As mention in \cite{2019MNRAS.486.5482G} those parameters do not impact significantly the identification of the center of the clumps. Since this procedure does not differentiate between dark matter, gas and stars, we use it as a preliminary step. Initially, we select a clump detected by PHEW located in the galactic disc and then we group the inner over-dense gas cells as our target cloud. This second step allows us to further focus on the star-forming cells inside each cloud. The star formation efficiency for the detected clouds is computed as
\begin{equation}\label{eq:obseff}
\tilde{\epsilon}_{\mathrm{ff}}=\frac{M_{\star,\mathrm{y}}}{M_{\mathrm{pre}}}\frac{t_{\mathrm{ff}}}{t_{\star,\mathrm{y}}}
\end{equation}

\begin{figure*}
  \centering
  \begin{subfigure}[b]{0.45\linewidth}
    \includegraphics[width=\linewidth]{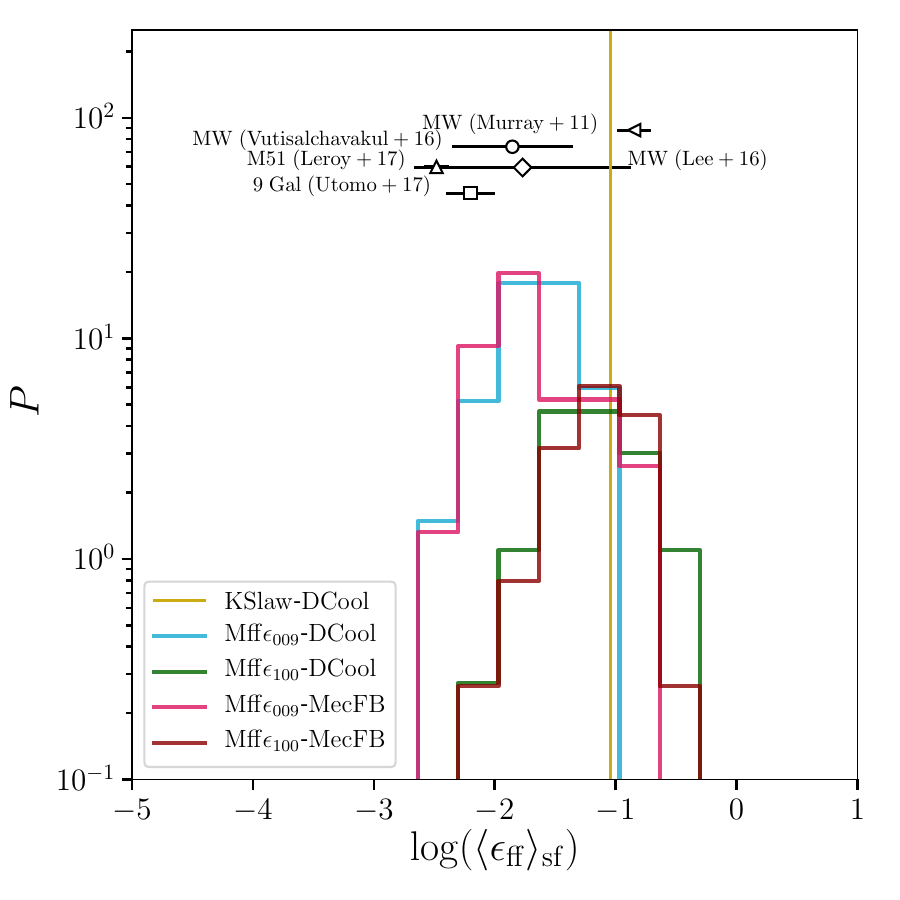}
    \caption{}
  \end{subfigure}
  \begin{subfigure}[b]{0.45\linewidth}
    \includegraphics[width=\linewidth]{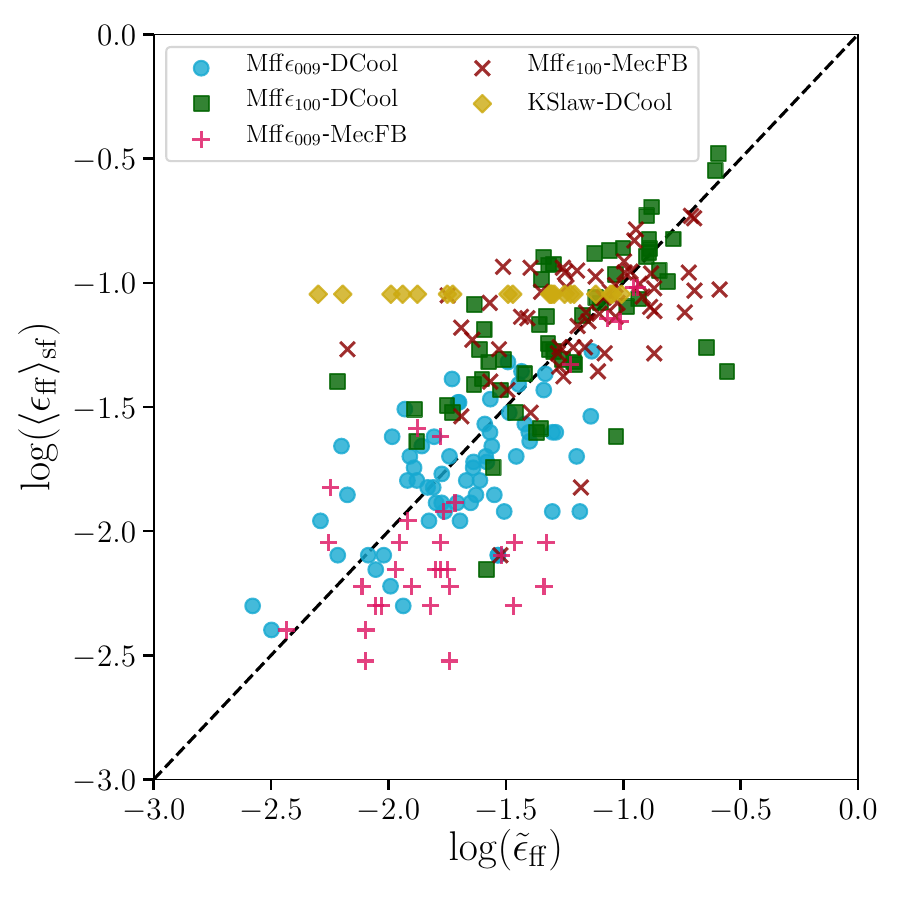}
    \caption{}
  \end{subfigure}
  
  \caption{(a) The star formation efficiency per free-fall time of the star-forming cell for all the runs compared with observations of giant molecular clouds (GMC) in the Milky Way \citep{2011ApJ...729..133M,2016ApJ...833..229L,2016ApJ...831...73V}, M51 \citep{2017ApJ...846...71L} and for a set of nine different galaxies \citep{2018ApJ...861L..18U}. (b) The correlation of the two ways of computing the efficiency of the GMC for the five runs, the observational efficiency, $\tilde{\epsilon}_{\mathrm{ff}}$, and the mean of the individual efficiencies of the star-forming cells inside the detected GMCs,$\langle\epsilon_{\mathrm{ff}}\rangle_{\mathrm{sf}}$.  }
\label{fig:efficiency}
\end{figure*}

\noindent where $M_{\star,\mathrm{y}}$ is the mass of the young stars (age $< t_{\star,\mathrm{y}}$) found inside the cloud, $M_{\mathrm{pre}}=M_{\mathrm{GMC}}+M_{\star,\mathrm{y}}$ is the gas mass of the GMC prior to the formation of the star and we take $t^{\star,\mathrm{y}}=4$ Myr which is consistent with \citet{2011ApJ...729..133M} and \citet{2019MNRAS.486.5482G}. The detected GMC in the simulation present comparable efficiencies but exceed the mass range of the observations, this effect was already observed by \cite{2018MNRAS.479.3167G} since GMC observations are done in the line of sight, i.e. in 2D while our reconstruction is in 3D. We do an extra comparison by only taking the star-forming cells inside each GMC and compare the mean of their individual SF efficiency $\langle\epsilon_{\mathrm{ff}}\rangle_{\mathrm{sf}}$ as calculated by the multi-ff model (see equation \ref{eq:efficiency}) and the sum of their masses $M_{\mathrm{SF_{cells}}}^{\mathrm{GMC}}$. This comparison is shown in the right panel of figure \ref{fig:effvsMass} for the five runs and results in better agreement with observations in both mass and efficiencies for most simulations. Particularly, ultra-efficient sites in the Mff$\epsilon_{009}$-MecFB run are shown here as the massive outliers, this cells are very dense and therefore turn more gas mass into stars.

 Furthermore, we look at the distribution of values of the mean individual efficiencies of the star-forming cells inside the detected GMC, $\langle\epsilon_{\mathrm{ff}}\rangle_{\mathrm{sf}}$, and compare them with different observations for the MW and M51 \cite{2014ApJ...782..114E,2016ApJ...833..229L,2016ApJ...831...73V,2017ApJ...846...71L,2018ApJ...861L..18U} as shown in the left panel of figure \ref{fig:efficiency}. The fixed $\epsilon_{\mathrm{ff}}$ of the KSlaw-DCool run is shown as a vertical line. Even if the observations in the MW from \cite{2011ApJ...729..133M} have a good agreement with the KSlaw-DCool, we observe general agreement with most observations for all the runs with multi-ff SF (Mff$\epsilon_{009}$-DCool, Mff$\epsilon_{009}$-MecFB, Mff$\epsilon_{100}$-DCool and Mff$\epsilon_{100}$-MecFB) where the star formation efficiency is computed directly from gas features. In the right panel of figure \ref{fig:efficiency}, we show the correlation between the two ways of calculating the efficiency of the GMC.For the KSlaw-DCool run, even if one can derive a range of values for $\tilde{\epsilon}_{\mathrm{ff}}$ following equation \ref{eq:obseff}, the actual efficiency plugged in the calculation inside the cell is constant (equation \ref{eq:starformation}) and tuned purposely to agree with observations. But, as this value is constant we cannot correlate it with $\tilde{\epsilon}_{\mathrm{ff}}$ as it is the case for the simulations with the multi-freefall star formation (by meaning the values of the actual star-forming cells of the identified molecular clouds) accounting for a more consistent and less  tuned scheme.
 

\section{Summary and conclusions}\label{sec:Conclusions}
We perform simulations of one selected spiral galaxy in a cosmological environment with the RAMSES code to explore the impact of sub-grid baryonic physics implementation. The galaxy labelled Mochima is chosen according to the host halo mass and the global stellar mass to be close to the MW values. We focus on star formation and SN feedback, as these are known to be two determining processes shaping galaxy formation and evolution. Starting from the same initial conditions, different implementations of the baryonic physics yield significant changes in the shape and properties of the final galaxy. We reach a resolution of 35 pc inside a cosmological box of 36 Mpc. All of the runs presented here exhibit a spiral disc at redshift 0 inside a Milky way size DM halo. This resemblance allows us to make comparisons of our simulations with observations of the Milky Way or local spiral galaxies which is done in two main blocks, comparing global properties and local properties of the galaxies.

Our strategy consists in starting with the popular sub-grid implementations used in such simulations, e.g. Schmidt law SF, which allows star formation in gas regions (cells) with densities above a certain threshold and with a fixed efficiency. Together with delayed cooling feedback, which consists of eliminating cooling temporarily in the expanding SN event. It is known that these models, while successful in describing large scale features of galaxy populations, lack details on the physical process they aim to represent. Therefore we depart from this ``control" run labelled KSlaw-DCool of the Mochima galaxy by changing one sub-grid recipe, namely, the star formation to get the second group of runs: multi-ff SF and delayed cooling with strong PSFB, Mff$\epsilon_{009}$-DCool, and weak PSFB, Mff$\epsilon_{100}$-DCool. In these cases, the efficiency of the star formation is no longer fixed, and it depends on the turbulence in the local gas. To this end, we have included a sub-grid model to propagate the turbulent kinetic energy of the gas through time. For the third group of runs, we use the multi-ff SF model together with a model of mechanical feedback where the main stages of the Sedov-Taylor explosion are considered with additional strong PSFB, Mff$\epsilon_{009}$-MecFB, and weak PSFB, Mff$\epsilon_{100}$-MecFB. Our main results are:

As mention before, we observe a spiral galaxy in all five runs, although with fairly different morphologies. The KSlaw-DCool results in a smooth distribution of stars with a few concentrated star formation sites in the disc and with most of its stars concentrated in the bulge. While once only the SF implementation is changed in the second group of runs, we observed a less extended disc but with better populated spiral arms. Here the strong PSFB results in one of our so-called successful galaxies, Mff$\epsilon_{009}$-DCool, with respect to the discussed tests. Between the second and third groups, the feedback implementation changed, and the mechanical feedback is introduced. In the disc morphology of the Mff$\epsilon_{009}$-MecFB case, we start observing that the combination of mechanical feedback implementation with a strong PSFB is not able to disrupt star-forming clouds. Several very bright spots of highly efficient star formation are observed in the disc together with an extremely bright bulge (figure \ref{fig:maps}). This situation is solved by factoring out the PSFB in the Mff$\epsilon_{100}$-MecFB case, here the resulting galaxy is much more smooth and better populated, hence is one of our two successful runs.

Globally the five runs present a good ratio between the stellar mass and the DM mass of the halo agreeing with abundance matching techniques and MW mass. In particular, we observe an excess population of satellite galaxies in the Mff$\epsilon_{009}$-MecFB run compared to the other two runs (figure \ref{fig:SHMR}). The Kennicutt-Schmid relation is reproduced well by the runs with delay cooling and the Mff$\epsilon_{100}$-MecFB run. On the other hand, the Mff$\epsilon_{009}$-MecFB run exhibits a very efficient gas consumption and does not reproduce the KS slope (figure \ref{fig:KSL}). In the case of the cosmological baryonic ratio and the Tully-Fisher relation, all galaxies are in good agreement with observations (figures \ref{fig:barfrac} and \ref{fig:TF}).

When it comes to star formation history, we observe the main difference between the different feedback combinations. The combination of mechanical feedback and strong PSFB (Mff$\epsilon_{009}$-MecFB) is not able to prevent star formation at very early stages of the galactic history at redshift 3-4, where it is allowing most of the mass of the galaxy to be formed. On the other hand, for $z<$1.5, we observe clearly the difference between the two star formation implementations. The four runs with multi-ff SF show similar SFR one order of magnitude above the SFR in the Schmidt law SF run (figure \ref{fig:SFRH}).

Locally, we study the agreement of the inner features of the galaxies with MW observations, starting with the rotation curves where we observe better agreement in the Mff$\epsilon_{009}$-DCool run. The stars in the KSlaw-DCool and Mff$\epsilon_{100}$-DCool run are rotating about 50 km/s slower than the stars of the MW for certain radii. The runs with mechanical feedback (Mff$\epsilon_{009}$-MecFB and Mff$\epsilon_{100}$-MecFB) present a diverging velocity profile in the centre due to the mass of the bulge (figure \ref{fig:RC}). Further comparisons with the MW stellar mass distribution in the disc were performed and show impressive agreement in the runs with delayed cooling and the Mff$\epsilon_{100}$-MecFB. Alternatively, the Mff$\epsilon_{009}$-MecFB run exceeds what is expected from the MW mass model.  

The resolution achieved in these simulations is comparable to recent observations of star-forming clouds in the MW and local spiral galaxies. We store the information of the gas in the star-forming cells during our simulations to study the environment that triggers star formation in our five runs. Furthermore, we compare these star-forming environments with molecular cloud observations using observables like density, temperature and efficiency per freefall time. Here we observe i) that the Schmidt law SF aside for having a fixed star formation efficiency which already disagrees with observations, forms stars in regions with higher temperatures than would be expected (figure \ref{fig:PhaseD}), ii) A high non-linearity in the galaxy evolution problem allows different combinations of feedback implementations to result in interesting galactic distribution as we observe for the cases of the Mff$\epsilon_{009}$-DCool and Mff$\epsilon_{100}$-MecFB runs but iii) other combinations can result in ultra-efficient star-forming sites (Mff$\epsilon_{009}$-MecFB) or very faint stellar disc populations (KSlaw-DCool and Mff$\epsilon_{100}$-DCool).

While it seems that by adding complexity to the sub-grid models we end up generating higher stellar masses, there is gain in morphological aspects, dynamical aspects of the overall galaxy and local star-forming gas features, depending on the FB combinations. At least in favour of the addition of turbulence to the star formation strategies. In the case of the not-successful FB combinations, possible reasons of the difference between observation and our results are i) unaccounted feedback physics such as radiation feedback, cosmic rays or even AGN feedback, that usually serve as a justification of the strength of the delayed cooling method. ii) The resolution reached in our simulations is still not enough for this implementation to affect the local environment of the SN explosion correctly and iii) following the lines of the last point we might be suffering from overcooling at galactic scales. Higher resolutions are still required.

Finally, on the combination of feedback implementations we attempt to bracket the possible values of the free parameter in the multi freefall star formation model, $\epsilon$, but conclude that its value also depends on the SN feedback recipe. For delayed cooling lower values of $\epsilon$ seem to be favoured, contrary to the mechanical feedback where higher values of $\epsilon$ are favoured. This last scenario is consistent with predicted values for epsilon in semi-analytic models, where $\epsilon= 0.3-0.7$ are suggested by \citet{2012ApJ...761..156F}.

Generally, our simulations exhibit an excess in early star formation generating a dense and massive bulge of old stars. The associated steep central gravitational potential certainly prevents the formation of bars. Such situation represents a common issue in similar high resolution cosmological simulations.

The present studies show the need for improved sub-grid implementations, in particular for the interplay between turbulence, star formation and supernova feedback in cosmological environments. This work also highlights the inner degeneracies of the galaxy formation problem. More precise diagnostics could discriminate amongst the different baryonic models.
\section*{Acknowledgements}

We thank Valentin Perret, Joakim Roshdahl, Benoit Famaey, Lorenzo Posti, Mihael Petac, Gary Mamon, Jean-Charles Lambert and Andr\'e Tilquin for fruitful discussions and support. This work was founded by OCEVU Labex (ANR-11-LABX-0060) and the A*MIDEX project (ANR-11-IDEX-0001-02) funded by the ``Investissements d'Avenir" French government program managed by the ANR.
Centre de Calcul Intensif d’Aix-Marseille is acknowledged for granting access to its high performance computing resources. This work benefited from the scientific environment of the French ANR
project GaDaMa (ANR-18-CE31-0006).

\section*{Data Availability}
The data underlying this article will be shared on reasonable request to the corresponding author.




\bibliographystyle{mnras}
\bibliography{fulllib} 








\label{lastpage}
\end{document}